\documentclass[aps,pra,preprint,superscriptaddress,showpacs]{revtex4-1}
\usepackage{bm}
\usepackage{amsmath}
\allowdisplaybreaks
\usepackage{dcolumn}
\newcolumntype{.}{D{x}{}{-1}}

\newcommand*{\centt}[1]{\multicolumn{1}{c}{#1}}
\newcolumntype{w}[1]{D{.}{.}{#1}}
\def\mynote1#1{{\color{blue}{\textsc{\footnotesize Question/Comment:} #1}}}
\def\mynote2#1{{\color{magenta}{#1}}}

\begin{document}

\preprint{Version 1.1}

\title{Higher-order recoil corrections for singlet states of the helium atom}
\author{Vojt\v{e}ch Patk\'o\v{s}}
\affiliation{Faculty of Physics, University of Warsaw,
             Pasteura 5, 02-093 Warsaw, Poland}

\author{Vladimir A. Yerokhin}
\affiliation{Center for Advanced Studies, Peter the Great St.~Petersburg Polytechnic University,
Polytekhnicheskaya 29, 195251 St.~Petersburg, Russia}

\author{Krzysztof Pachucki}
\affiliation{Faculty of Physics, University of Warsaw,
             Pasteura 5, 02-093 Warsaw, Poland}

\date{\today}

\begin{abstract}

We investigate the finite nuclear mass corrections in the helium atom in order to resolve a
significant disagreement between the $2^3S - 2^3P$ and $2^3S - 2^1S$ transition isotope shifts.
These two transitions lead to discrepant results for the nuclear charge radii difference between
$^4$He and $^3$He. The accurate treatment of the finite nuclear mass effects is quite complicated and 
requires the use of the quantum field theoretical approach. We derive the $\alpha^6\,m^2/M$
correction with the help of nonrelativistic QED and dimensional regularization of the three body
Coulombic system, and present accurate numerical results for low-lying states. The previously
reported $4\,\sigma$ discrepancy in the nuclear charge radius difference between $^3$He and
$^4$He from two different atomic isotope shift transitions is confirmed, which calls for
verification of experimental transition frequencies.

\end{abstract}

\maketitle

\section{Introduction}
The atomic spectroscopy of light atoms has reached the level of precision that allows the
determination of nuclear parameters from measured transition frequencies, in particular the nuclear
charge radius. The best known example is the hydrogen spectroscopy from which one obtains the
proton mean square charge radius of $r_p = 0.8758(77)$ fm, in agreement with the result derived
from the electron-proton elastic scattering, $0.895(18)$ fm \cite{mohr:16:codata}. Both these
values are in significant disagreement with the result derived from the muonic hydrogen Lamb
shift, $r_p = 0.84087(39)$ fm \cite{pohl:10, antognini:13}. This discrepancy attracted much
attention from the scientific community and became known as the proton charge radius puzzle
\cite{review}. Up to now the determination of nuclear charge radii from light atoms other than
hydrogen has been limited by the lack of sufficiently accurate theory. It was only possible to find
the nuclear charge radii differences from the isotope shifts of atomic transition frequencies
\cite{rmp_review}. Bearing in mind the discrepancy between the electronic and the muonic hydrogen
determinations of the proton charge radius, we investigate the isotopic differences in the nuclear
charge radii in order to explore other potential discrepancies. Indeed, the nuclear charge radii
difference $\delta r^2$ between $^4$He and $^3$He was determined to be $1.069(3)$ $\textrm{fm}^2$
from the $2^3S - 2^3P$ transition \cite{herecoil} and $1.027(11)$ $\textrm{fm}^2$ from the $2^3S -
2^1S$ transition \cite{heis}. The $4\,\sigma$ discrepancy between these two results could be
explained by a $8.8$ kHz shift in the $2^3S - 2^1S$ transition, a small correction which in
principle might have been overlooked in previous calculations. The corresponding shift in the $2^3S
- 2^3P$ transition would have to be much larger, $49.7$ kHz, and thus is less probable. In this
work we calculate the last unknown correction, of order $\alpha^6\,m^2/M$, which might contribute
at this level of accuracy. We find out that the result for the isotope shift of the $2^3S - 2^1S$
transition almost coincides with our previous estimate \cite{heis}, namely $2.73$~kHz versus
$2.75(69)$~kHz. Since we do not see any possibility to miss a 8.8 kHz effect in our theoretical
predictions, we are in a position to claim a discrepancy between the isotope shift in the $2^3S -
2^3P$ \cite{cancio:04, cancio:12:3he,shiner:95} and $2^3S - 2^1S$ \cite{rooij:11} transition
frequencies.

\section{Notations}
In this work we closely follow our previous paper devoted to nuclear recoil effects for triplet
states of helium \cite{herecoil} and use the same notations. The reader may consider checking that
paper first, but nevertheless we repeat here the main principles. The operators, energies, and wave
functions for a nucleus with a finite mass $M$ are marked with indices ``$M$'': $X_M$, $E_M$, and
$\phi_M$. The operators, energies, and wave functions in the infinite nuclear mass limit are
without indices: $X$, $E$, and $\phi$. The recoil corrections to the operators and energies are denoted
by $\delta_M X$ and $\delta_M E$,
\begin{equation}\label{01}
X_M\equiv X +\frac{m}{M}\delta_M X+O\Bigl(\frac{m}{M}\Bigr)^2\,,
\end{equation}
\begin{eqnarray}\label{02}
E_M=E+\frac{m}{M}\,\delta_M E+O\Bigl(\frac{m}{M}\Bigr)^2\,.
\end{eqnarray}
We also introduce the shorthand notations:
\begin{equation}\label{03}
\langle X\rangle_M\equiv\langle\phi_M|X|\phi_M\rangle\,,
\end{equation}
and
\begin{eqnarray}\label{04}
\delta_M\,\langle X\rangle
&\equiv&
\biggl\langle\phi\,\biggl|\frac{ \vec{P}_I^2}{2}\frac{1}{(E-H)'}\,X\,\biggr|\,\phi\biggr\rangle
+\biggl\langle\phi\,\biggl|\, X\,\frac{1}{(E-H)'}\,\frac{ \vec{P}_I^2}{2}\biggr|\,\phi\biggr\rangle\,,
\end{eqnarray}
where $\vec{P_I}$ is the momentum of the nucleus in the center-of-mass frame, and
$H$, $E$, and $\phi$  are the nonrelativistic
Hamiltonian, energy, and wave function in the infinite nuclear mass limit.

According to the QED theory, the expansion of energy levels in powers of $\alpha$ has the form
\begin{eqnarray}\label{05}
E_M\Bigl(\alpha,\frac{m}{M}\Bigr)=E_M^{(2)}+E_M^{(4)}+E_M^{(5)}+E_M^{(6)}+E_M^{(7)}+O(\alpha^8),
\end{eqnarray}
where $E_M^{(n)}$ is a contribution of order $m\,\alpha^n$ and may include powers of $\ln\alpha$.
$E_M^{(n)}$ is in turn expanded in powers of the electron-to-nucleus mass ratio $m/M$
\begin{eqnarray}\label{06}
E_M^{(n)}=E^{(n)}+\frac{m}{M}\,\delta_M E^{(n)}+O\Bigl(\frac{m}{M}\Bigr)^2.
\end{eqnarray}
We are interested here in $E^{(6)}_M$, which can be expressed as
\begin{eqnarray}\label{07}
E_M^{(6)}=\left\langle H_M^{(4)}\,\frac{1}{(E_M-H_M)'}\,H_M^{(4)}\right\rangle_M
+ \bigl\langle H_M^{(6)}\bigr\rangle_M = A_M + B_M\,,
\end{eqnarray}
where the last equation is a definition of $A_M$ and $B_M$.
In this paper we derive the recoil part of this correction $\delta_M E^{(6)}$ for singlet states
in helium. The computational approach is similar to the one used for triplet states in Ref.  \cite{herecoil}
and to the nonrecoil $\alpha^6\,m$ correction for singlet states in Ref. \cite{pachucki:06hesinglet}.

\section{Dimensional regularization}
Since individual terms in $E^{(6)}$ are divergent they have to be regularized.
We found in Ref. \cite{pachucki:06hesinglet} that the most convenient regularization
is the dimensional one, although it seems to be very exotic for atomic systems.
In this regularization, the dimension of space is assumed to be $d=3-2\,\epsilon$.
The photon propagator, and thus the Coulomb interaction preserves its form in the momentum representation,
while in the coordinate representation the Coulomb potential is
\begin{eqnarray}
\label{08}
\int\frac{d^dk}{(2\pi)^d}\,\frac{4\pi}{k^2}\,e^{i\vec{k}\cdot\vec{r}} = \pi^{\epsilon-1/2}\,\Gamma(1/2-\epsilon)\, r^{2\epsilon-1}
\equiv \frac{C_1}{r^{1-2\epsilon}}.
\end{eqnarray}
The elimination of singularities is performed in atomic units
by the transformation
\begin{eqnarray}\label{09}
\vec{r}\rightarrow(m\alpha)^{-1/(1+2\epsilon)}\,\vec{r}
\end{eqnarray}
and pulling common factors $m^{(1-2\epsilon)/(1+2\epsilon)}\,\alpha^{2/(1+2\epsilon)}$
and $m^{(1-10\epsilon)/(1+2\epsilon)}\,\alpha^{6/(1+2\epsilon)}$
from $H$ and $H^{(6)}$, respectively.
The nonrelativistic Hamiltonian of hydrogen-like systems is
\begin{eqnarray}\label{10}
H=\frac{\vec{p}\,^2}{2}-Z\frac{C_1}{r^{1-2\epsilon}}\,,
\end{eqnarray}
and that of helium-like systems is
\begin{equation}\label{11}
H = \frac{\vec{p_1}^2}{2}+\frac{\vec{p_2}^2}{2}+V  \,,
\end{equation}
where
\begin{equation}\label{12}
V = -Z\frac{C_1}{r_1^{1-2\epsilon}}-Z\frac{C_1}{r_2^{1-2\epsilon}}+\frac{C_1}{r_{12}^{1-2\epsilon}}
  \equiv \biggl[-\frac{Z}{r_1} -\frac{Z}{r_2} + \frac{1}{r}\biggr]_\epsilon\,.
\end{equation}
We calculate further integrals involving the photon propagator in the Coulomb gauge as follows
\begin{eqnarray}\label{13}
\int\frac{d^d k}{(2\pi)^d}\,\frac{4\pi}{k^4}\,
\left(\delta^{ij}-\frac{k^i\,k^j}{k^2}\right)\,e^{i\vec{k}\cdot\vec{r}}
&=&\pi^{\epsilon-1/2}\,r^{-1+2\epsilon}\left[\frac{3}{16}\,\delta^{ij}\,
\Gamma(-1/2-\epsilon)\,r^2+\frac{1}{8}\,\Gamma(1/2-\epsilon)\,r^i\, r^j\right]\nonumber\\
&\equiv&\frac{1}{8}\,\left[\frac{r^i\,r^j}{r}-3\,\delta^{ij}\,r\right]_\epsilon\,,
\end{eqnarray}
and
\begin{eqnarray}\label{14}
\int \frac{d^d k}{(2\pi)^d}\,\frac{4\pi}{k^2}\,\left(\delta^{ij}-\frac{k^i\,k^j}{k^2}\right)\,e^{i\vec{k}\cdot\vec{r}}
&=&\pi^{\epsilon-1/2}\,r^{-3+2\epsilon}\left[\frac{1}{2}\delta^{ij}\,\Gamma(1/2-\epsilon)\,r^2+\Gamma(3/2-\epsilon)\,r^i\,r^j\right]\nonumber\\
&\equiv&\frac{1}{2}\,\biggl[\frac{r^i\,r^j}{r^3} + \frac{\delta^{ij}}{r}\biggr]_\epsilon .
\end{eqnarray}
The solution of the stationary Schr\"{o}dinger equation $H\,\phi=E\,\phi$ is denoted by $\phi$,
and we will never need its explicit (and unknown) form in $d$ dimensions.

\section{Effective Hamiltonian in $d-$dimensions}
We pass now to the effective Hamiltonian terms in Eq. (\ref{07}).
The Breit-Pauli Hamiltonian $H_M^{(4)}$ \cite{bs, pachucki:06hesinglet} is split into two parts (with $r_{12}\equiv r$,
$r_{aI}\equiv r_a$ and $\vec{P}\equiv\vec{p}_1+\vec{p}_2$)
\begin{equation}\label{15}
H_M^{(4)} = H_A^M+H_C^M\,,
\end{equation}
where
\begin{eqnarray}\label{16}
H_A^M & = & -\frac{1}{8}\,(p_1^4+p_2^4)+
\frac{Z\,\pi}{2}\,[\delta^d(r_1)+\delta^d(r_2)]+\pi\,(d-2)\,\delta^d(r)
-\frac{1}{2}\,p_1^i\,
\biggl[\frac{\delta^{ij}}{r}+\frac{r^i\,r^j}{r^3}\biggr]_\epsilon\,p_2^j \nonumber\\
&&-\frac{Z}{2}\,\frac{m}{M}\,\biggl[
p_1^i\,\biggl[\frac{\delta^{ij}}{r_1} + \frac{r_1^i\,r_1^j}{r_1^3}\biggr]_\epsilon+
p_2^i\,\biggl[\frac{\delta^{ij}}{r_2} + \frac{r_2^i\,r_2^j}{r_2^3}\biggr]_\epsilon\biggr]\,P^j\,,
\end{eqnarray}
and
\begin{eqnarray}\label{17}
H_C^M& = & \left[
\frac{Z}{4}\biggl(
\frac{\vec{ r}_1}{r_1^3}\times\vec{ p}_1-
\frac{\vec{ r}_2}{r_2^3}\times\vec{ p}_2\biggr)+
\frac{1}{4}\,\frac{\vec{ r}}{r^3}\times
(\vec{ p}_1+\vec{ p}_2)
+\frac{Z}{2}\,\frac{m}{M}\,\biggl(
\frac{\vec r_1}{r_1^3} - \frac{\vec r_2}{r_2^3}\biggr)\times\vec{P}
\right]\,\frac{\vec{\sigma}_1-\vec{\sigma}_2}{2}\,.
\end{eqnarray}
$H_C^M$ in the above equation can be represented in $d=3$ as it does not lead to any singularities.
The other terms in $H_M^{(4)}$ do not contribute to energies of singlet states.
The corresponding second-order correction is
\begin{equation}\label{18}
A_M = \langle H_A^M\,\frac{1}{(E_M-H_M)'}\,H_A^M\rangle_M +
      \langle H_C^M\,\frac{1}{(E_M-H_M)'}\,H_C^M\rangle_M\,,
\end{equation}
whereas the first-order contribution is given by
\begin{equation}\label{19}
B_M= \langle \sum_{i=1,12} H_i^M\rangle_M
\end{equation}
where, following Ref. \cite{herecoil} $H_i^M$ in arbitrary $d-$dimensions are as follows
\begin{equation}\label{20}
H_1^M =  \frac{p_1^6}{16}+\frac{p_2^2}{16}\,,
\end{equation}
\begin{equation}\label{21}
H_2^M =
\frac{(\nabla_1 V)^2+(\nabla_2 V)^2}{8}
+\frac{5}{128}\,\biggl(\,[p_1^2,[p_1^2,V]]+[p_2^2,[p_2^2,V]]\biggr)
-\frac{3}{64}\,\,\biggl(\Bigl\{p_1^2\,,\,\nabla_1^2 V\Bigr\}+
\Bigl\{p_2^2\,,\,\nabla_2^2 V\Bigr\}\biggr)\,,
\end{equation}
\begin{equation}\label{22}
H_{3}^M = \frac{1}{64}\biggl(
-4\,\pi\,\nabla^2\delta^3(r)+\frac{16\,\pi}{d\,(d-1)}\,\sigma^{kl}_1\sigma^{kl}_2\,p_1^i\,
\biggl[\frac{2}{3}\,\delta^{ij}\,4\,\pi\,\delta^3(r)+\frac{1}{r^5}\,(3\,r^i\,r^j-\delta^{ij}\,r^2)\biggr]_\epsilon\,p_2^j\biggr)\,,\\
\end{equation}
\begin{eqnarray}\label{23}
H_4^M &=&
\frac{1}{2}\,\bigl(p_1^2+p_2^2\bigr)\,p_1^i\,\biggl[\frac{1}{2\,r}\left(\delta^{ij}+\frac{r^ir^j}{r^2}\right)\biggr]_\epsilon p_2^j
+\frac{(p_1^2+p_2^2)}{8}\,\frac{\sigma_1\,\sigma_2}{d}\,4\,\pi\,\delta^3(r)\nonumber\\
&&+\,\frac{Z}{2\,M}\,\biggl(p_1^2\,p_1^i\biggl[\frac{1}{2\,r_1}\biggl(\delta^{ij}+\frac{r_1^ir_1^j}{r_1^2}\biggr)\biggr]_\epsilon\,P^j
+p_2^2\,p_2^i\biggl[\frac{1}{2\,r}\biggl(\delta^{ij}+\frac{r_2^ir_2^j}{r_2^2}\biggr)\biggr]_\epsilon\,P^j\biggr)\,,\nonumber\\
\end{eqnarray}
\begin{equation}\label{24}
H_5^M =  \frac{\sigma^{ij}_1\,\sigma^{ij}_2}{2\,d}\,\biggl(
-\frac{1}{2}\,\biggl[\frac{\vec r}{r^3}\biggr]_\epsilon\,(\nabla_1 V +\nabla_2 V) +
\frac{1}{16}\,\biggl(\biggl[\biggl[\biggl[\frac{1}{r}\biggr]_\epsilon\,,\,p_1^2\biggr]\,,\,p_1^2\biggr]
+\biggl[\biggl[\biggl[\frac{1}{r}\biggr]_\epsilon\,,\,p_2^2\biggr]\,,\,p_2^2\biggr]\biggr)\biggr)\,, \\
\end{equation}
\begin{eqnarray}\label{25}
H_6^M &=&
\frac{1}{8}\,p_1^i\,\frac{1}{r^2}\biggl(\delta^{ij}+3\,\frac{r^ir^j}{r^2}\biggr)\,p_1^j
+\frac{1}{8}\,p_2^i\,\frac{1}{r^2}\biggl(\delta^{ij}+3\,\frac{r^ir^j}{r^2}\biggr)\,p_2^j+\frac{(d-1)}{4}\biggl[\frac{1}{r^4}\biggr]_\epsilon
\nonumber \\ &&
+\,\frac{Z}{4}\frac{m}{M}\biggl[\,p_2^i\,\biggl(\frac{\delta^{ij}}{r}+\frac{r^i\,r^j}{r^3}\biggr)\,
\biggl(\frac{\delta^{jk}}{r_1}+\frac{r_1^j\,r_1^k}{r_1^3}\biggr)\,P^k+(1\leftrightarrow2)\biggr]
+\frac{Z^2}{8}\,\frac{m}{M}\nonumber \\ &&
\times\biggl[p_1^i\,\biggl(\frac{\delta^{ij}}{r_1}+3\,\frac{r_1^i\,r_1^j}{r_1^3}\biggr)\,\,p_1^k
+p_2^i\,\biggl(\frac{\delta^{ij}}{r_2}+3\,\frac{r_2^i\,r_2^j}{r_2^3}\biggr)\,\,p_2^k
+2\,p_1^i\,\biggl(\frac{\delta^{ij}}{r_1}+\frac{r_1^i\,r_1^j}{r^3}\biggr)\,
\biggl(\frac{\delta^{jk}}{r_2}+\frac{r_2^j\,r_2^k}{r_2^3}\biggr)\,p_2^k\nonumber\\
&&+\,\frac{\sigma^{ij}_1\,\sigma^{ij}_1}{d}\,\biggl[\frac{1}{r_1^4}\biggr]_\epsilon
+\frac{\sigma^{ij}_2\,\sigma^{ij}_2}{d}\,\biggl[\frac{1}{r_2^4}\biggr]_\epsilon
+2\,\frac{\sigma^{ij}_1\,\sigma^{ij}_2}{d}\,\frac{\vec r_1}{r_1^3}\,\frac{\vec r_2}{r_2^3}\,\biggr]\,,
\end{eqnarray}
\begin{eqnarray}\label{26}
H_{7a}^M &=& -\frac{1}{8}\,\biggl\{
\bigl[p_1^i,V\bigr]\,\frac{r^i\,r^j-3\,\delta^{ij}\,r^2}{r}\,
\bigl[V,p_2^j\bigr]
+\bigl[p_1^i,V\bigr]\,\biggl[\frac{p_2^2}{2},
 \frac{r^i\,r^j-3\,\delta^{ij}\,r^2}{r}\biggr]\,p_2^j \nonumber \\ &&
+p_1^i\,\biggl[\frac{r^i\,r^j-3\,\delta^{ij}\,r^2}{r},
\frac{p_1^2}{2}\biggr]\,\bigl[V,p_2^j\bigr]
+p_1^i\,\biggl[\frac{p_2^2}{2},\biggl[
\frac{r^i\,r^j-3\,\delta^{ij}\,r^2}{r},
\frac{p_1^2}{2}\biggr]\biggr]\, p_2^j\biggr\}\,,
\end{eqnarray}
\begin{equation}\label{27}
H_{7c}^M =\frac{\sigma^{ij}_1\,\sigma^{ij}_2}{16\,d}\,
\biggl[p_1^2,\biggl[p_2^2,\biggl[\frac{1}{r}\biggr]_\epsilon\biggr]\biggr] \,, \\
\end{equation}
\begin{equation}\label{28}
H_{7d}^M = \frac{i\,Z}{8}\frac{m}{M}\,(\nabla_1^i V+\nabla_2^i V)\,\biggl(\biggl[H-E\,,\,
\frac{r_1^i\,r_1^j-3\,\delta^{ij}\,r_1^2}{r_1}\,p_1^j\biggr]+\biggl[H-E\,,\,
\frac{r_2^i\,r_2^j-3\,\delta^{ij}\,r_2^2}{r_2}\,p_2^j\biggr]\biggr)
\,.
\end{equation}
$H_{7b}^M$ would contain the spin-orbit type of interaction, but it vanishes for singlet states.
Further terms come from high energy photons and are known as pure, radiative and radiative recoil corrections,
which are the same as in hydrogenic systems \cite{eides}:
\begin{equation}\label{29}
H_{8}^M =  Z^3\,\frac{m}{M}\,\biggl(4\,\ln2 -\frac{7}{2}\biggr)\,\bigl[\delta^3(r_1)+\delta^3(r_2)\bigr]\,,
\end{equation}
\begin{equation}\label{30}
H_{9}^M = Z^2\,\frac{m}{M}\,\left(\frac{35}{36} - \frac{448}{27 \pi^2} - 2 \ln(2)
            +\frac{6 \zeta(3)}{\pi^2}\right)\,\bigl[\delta^3(r_1)+\delta^3(r_2)\bigr]\,,
\end{equation}
\begin{eqnarray}\label{31}
H_{10}^M &=& \pi\,Z^2\,\biggl(\frac{427}{96}- 2\,\ln(2) \biggr)\,
              \bigl[\delta^3(r_1)+\delta^3(r_2)\bigr]\nonumber\\
         && +\,\pi\,\biggl(\frac{6\zeta(3)}{\pi^2}-\frac{697}{27\pi^2}-8\ln(2)+\frac{1099}{72}\biggr)\,\delta^3(r)\,, \label{HRC1}
\end{eqnarray}
\begin{eqnarray}\label{32}
H_{11}^M &=&  \pi\,Z\,\biggl(-\frac{2179}{648\pi^2} -\frac{10}{27} +
              \frac{3}{2}\,\ln(2) -\frac{9\zeta(3)}{4\pi^2}\biggr)\,
              \bigl[\delta^3(r_1)+\delta^3(r_2)\bigr]\nonumber\\
         && + \,\pi\,\biggl(\frac{15\zeta(3)}{2\pi^2}+\frac{631}{54\pi^2}-5\ln(2)+\frac{29}{27}\biggl)\,\delta^3(r)\,.
\end{eqnarray}
The last term comes from the hard three-photon exchange between electrons. It was
originally calculated for positronium in Ref. \cite{czarnecki},
and for electrons its sign is reversed, see $H_H$ in Ref. \cite{pachucki:06hesinglet}
\begin{equation}
\label{h12}
H_{12}^M = \biggl(-\frac{1}{\epsilon}-4\ln \alpha
-\frac{39\,\zeta(3)}{\pi^2}+\frac{32}{\pi^2}-6\ln(2)+\frac{7}{3}\biggr)\,\frac{\pi\,\delta^d(r)}{4}\,,
\end{equation}
where by convention we pull out the common factor $\bigl[(4\pi)^\epsilon\,\Gamma(1+\epsilon)\bigr]^2$ from all
matrix elements.

\section{Elimination of singularities}
The principal problem of this approach is that both the first-order and the second-order
contributions in Eq. (\ref{07}) are divergent and the divergence cancels out only in the sum. To
achieve the explicit cancellation of the divergences, we (i) regularize the divergent contributions by
applying dimensional regularization with $d = 3-2\,\epsilon$, (ii) move singularities from the second-order
contributions to the first-order ones, and (iii) cancel algebraically the $1/\epsilon$ terms.

In the following we first consider the recoil correction coming from the second-order matrix elements,
i.e. the first term in Eq.~(\ref{07}), which is denoted by $A_M$. The recoil correction from the
second term in Eq.~(\ref{07}), denoted by $B_M$, is examined next.
It is the second-order contribution due to $H_A^M$ which is divergent and
therefore is treated in $d$ dimensions. To pull out divergences we rewrite $H_A^M$ as
\begin{equation}\label{33}
H_A^M = H_{R}^M + \bigl\{H_M-E_M,\,Q_M\bigr\}\,,
\end{equation}
where $Q_M = Q + \delta_M Q$ and
\begin{eqnarray}\label{34}
Q &=& -\frac14\biggl[\frac{Z}{r_1}+\frac{Z}{r_2}\biggr]_\epsilon+\frac{(d-1)}{4}\biggl[\frac{1}{r}\biggr]_\epsilon,\\\label{35}
\delta_M Q &=& \frac34\biggl[\frac{Z}{r_1}+\frac{Z}{r_2}\biggr]_\epsilon\,.
\end{eqnarray}
The operator $Q_M$ is the same as that in Ref. \cite{pachucki:06hesinglet} with the exception that it also
includes the recoil part $\delta_M Q$. The regular part of operator $H_A^M$ can be evaluated
in three dimensions to yield
\begin{equation}\label{36}
H_{R}^M  = H_{R} + \frac{m}{M}\delta_M H_{R}\,,
\end{equation}
\begin{equation}\label{37}
H_{R}\,|\phi\rangle = \biggl\{- \frac{1}{2}(E-V)^2
-\frac{Z}{4}\frac{\vec{r}_1\cdot\vec{\nabla}_1}{r_1^3}
-\frac{Z}{4}\frac{\vec{r}_2\cdot\vec{\nabla}_2}{r_2^3}
+\frac14\,\nabla_1^2\,\nabla_2^2
-p_1^i\,\frac{1}{2\,r}\biggl(\delta^{ij}+\frac{r^ir^j}{r^2}\biggr)\,p_2^j\biggr\}|\phi\rangle \,,
\end{equation}
\begin{eqnarray}\label{38}
\delta_M H_{R}\,|\phi\rangle &=& \biggl\{(E-V)\biggl(\frac{\vec{P}^2}{2}-\biggl\langle
  \frac{\vec{P}^2}{2} \biggr\rangle\biggr)
+ \frac{3Z}{4}\frac{\vec{r}_1\cdot\vec{\nabla}_2}{r_1^3}
+ \frac{3Z}{4}\frac{\vec{r}_2\cdot\vec{\nabla}_1}{r_2^3}
 \nonumber \\ &&
- \frac{Z}{2}\,p_1^i\,\frac{1}{r_1}\biggl(\delta^{ij}+\frac{r_1^i\,r_1^j}{r_1^2}\biggr)\,P^j
- \frac{Z}{2}\,p_2^i\,\frac{1}{r_2}\biggl(\delta^{ij}+\frac{r_2^i\,r_2^j}{r_2^2}\biggr)\,P^j\biggr\}|\phi\rangle\,,
\end{eqnarray}
where
\begin{equation}\label{39}
V = -\frac{Z}{r_1}-\frac{Z}{r_2} + \frac{1}{r},
\end{equation}
and the kinetic energy of the nucleus is $\langle\vec{P}^2/2\rangle=\delta_M E$. After
the transformation in Eq.~(\ref{33}) $A_M$ takes the form
\begin{eqnarray}\label{41}
A_M &=&
\sum_{a=R,C}\biggl\langle H_a^M\,\frac{1}{(E_M-H_M)'}\,H_a^M\biggr\rangle_M\nonumber\\
&&+\,\bigl\langle Q_M\,(H_M-E_M)\,Q_M\bigr\rangle_M
+2\,E_M^{(4)}\,\bigl\langle Q_M\bigr\rangle_M
-2\,\bigl\langle H_M^{(4)}\,Q_M\bigr\rangle_M \nonumber\\
&=& A_1^M+A_2^M \,,
\end{eqnarray}
where $A_1^M$ stands for the first term (i.e. the second-order contribution), and $A_2^M$
incorporates the remaining first-order matrix elements. Recoil corrections are obtained by
perturbing the second-order matrix element by the kinetic energy of the nucleus.
As a result $\delta_M A_1$ becomes
\begin{eqnarray}\label{42}
&&\delta_M A_1=
\sum_{a=R,C}\biggl\langle H_a\frac{1}{(E-H)'}\,\biggl[\frac{ \vec{P}^2}{2}-\delta_M E\biggr]\,\frac{1}{(E-H)'}\,H_a\biggr\rangle\nonumber\\
&&+\,2\,\biggl\langle H_a\,\frac{1}{(E-H)'}\,[\,H_a-\langle H_a\rangle\,]\,\frac{1}{(E-H)'}\,\frac{\vec{P}^2}{2}\biggr\rangle
+\,2\,\biggl\langle \delta_M H_a\frac{1}{(E-H)'}H_a\biggr\rangle,
\end{eqnarray}
while the first-order terms are
\begin{eqnarray}\label{43}
A_2^M&=&\langle Q\,(H_M-E_M)\,Q\rangle_M + 2\,E_M^{(4)}\langle Q\rangle_M - 2\,\langle H^{(4)}_M\,Q\rangle_M\nonumber\\
&&+\,\frac{m}{M}\,\biggl\{2\,\langle Q\,(H-E)\,\delta_M Q\rangle+2E^{(4)}\langle\delta_M Q\rangle-2\,\langle H_A\,\delta_M Q\rangle\biggr\}\,.
\end{eqnarray}
Reduction of these terms will be left to the Appendix A, and we present here
the final result for the recoil part
\begin{eqnarray}\label{44}
\delta_M A_2&=&
\delta_M\,\biggl\langle-\frac{3}{32}\biggl[\frac{Z^2}{r_1^4}+\frac{Z^2}{r_2^4}\biggr]_\epsilon+\frac{(d-1)(d-5)}{16}\biggl[\frac{1}{r^4}\biggr]_\epsilon
+\frac14\biggl(\frac{Z\vec{r}_1}{r_1^3}-\frac{Z\vec{r}_2}{r_2^3}\biggr)\cdot\frac{\vec{r}}{r^3}+2\,E^{(4)}\,Q\nonumber\\
&&+\,\frac{Z(Z-2)}{4}\pi\,\biggl(\frac{\delta^3(r_1)}{r_2}+\frac{\delta^3(r_2)}{r_1}\biggr)
-\frac14\,p_1^i\biggl(\frac{Z}{r_1}+\frac{Z}{r_2}-\frac{2}{r}\biggr)\frac{1}{r}\biggl(\delta^{ij}+\frac{r^ir^j}{r^2}\biggr)\,p_2^j\nonumber\\
&&+\,\frac{(d-1)}{4}\,\biggl[p_1^i,\biggl[p_2^j,\biggl[\frac{1}{r}\biggr]_\epsilon\biggr]\biggr]
\,\biggl[\frac{1}{2\,r}\biggl(\delta^{ij}+\frac{r^ir^j}{r^2}\biggr)\biggr]_\epsilon
+(E-V)^2\,Q+\frac18\,p_1^2\biggl(\frac{Z}{r_1}+\frac{Z}{r_2}\biggr)\,p_2^2\nonumber\\
&&-\,\frac{(d-1)}{8}\,p_1^2\,\biggl[\frac{1}{r}\biggr]_\epsilon\,p_2^2-\frac{(d-1)}{16}[p_1^2,[p_2^2,V]]
+\frac{Z\,\pi}{2}\biggl(\frac{\delta^3(r)}{r_1}+\frac{\delta^3(r)}{r_2}\biggr)\biggr\rangle
+\delta_M E^{(4)}\biggl(E+\biggl\langle\frac{1}{2r}\biggr\rangle\biggr)\nonumber\\
&&+\,\biggl\langle\frac{11}{32}\biggl[\frac{Z^2}{r_1^4}+\frac{Z^2}{r_2^4}\biggr]_\epsilon-\frac{3}{16}\frac{Z^2\,\vec{r}_1\cdot\vec{r}_2}{r_1^3r_2^3}
+\frac32\frac{E^{(4)}}{r}-3\,E E^{(4)}+\frac34\,(E-V)^2\biggl[\frac{Z}{r_1}+\frac{Z}{r_2}\biggr]_\epsilon\nonumber\\
&&-\,\frac38\,p_1^2\biggl(\frac{Z}{r_1}+\frac{Z}{r_2}\biggr)\,p_2^2+\frac34\,p_1^i\,\biggl(\frac{Z}{r_1}+\frac{Z}{r_2}\biggr)
\frac{1}{r}\biggl(\delta^{ij}+\frac{r^ir^j}{r^2}\biggr)\,p_2^j+2\,\delta_M E\,(E-V)\,Q\nonumber\\
&&+\,\frac{\pi\,Z}{4}\delta^3(r_1)\,\biggl(\frac{Z-6}{r_2}+2\,E+2\,Z^2\biggr)
+\frac{\pi\,Z}{4}\delta^3(r_2)\,\biggl(\frac{Z-6}{r_1}+2\,E+2\,Z^2\biggr)
\nonumber\\
&&+\,\vec{P}\biggl[\frac{E}{4}\biggl(\frac{Z}{r_1}+\frac{Z}{r_2}\biggr)
-\frac{E}{2r}+\frac14\biggl(\frac{Z}{r_1}+\frac{Z}{r_2}\biggr)^2
-\frac{3}{4r}\biggl(\frac{Z}{r_1}+\frac{Z}{r_2}\biggr)+\frac{1}{2r^2}\biggr]\vec{P}\nonumber\\
&&-\,\frac{Z}{4}\biggl[ P^i\left(\frac{\delta^{ij}}{r_1}+\frac{r_1^ir_1^j}{r_1^3}\right)\left(\frac{Z}{r_1}+\frac{Z}{r_2}-\frac{2}{r}\right)p_1^j+(1\leftrightarrow2)\biggr]
-\frac32 \pi\,Z\biggl(\frac{\delta^3(r)}{r_1}+\frac{\delta^3(r)}{r_2}\biggr)\biggr\rangle\,.
\end{eqnarray}

We examine now the recoil correction coming from $B_M$ in Eq. (\ref{19}).
For each of the operators $H_i^M=H_i+\frac{m}{M}\,\delta_M H_i$, the recoil correction is the sum
of two parts: (i) the perturbation of the nonrelativistic wave function, of $E$ and $H$ by the nuclear
kinetic energy in the nonrecoil part, and (ii) the expectation value of the recoil part
$\delta_M H_i$ (if present). The derivation is straightforward but tedious, therefore we have moved
its description to Appendix B and present here only the final result for the recoil correction
\begin{eqnarray}\label{46}
\delta_M B&=&\delta_M\,\biggl\langle\frac{7}{32}\biggl[\frac{Z^2}{r_1^4}+\frac{Z^2}{r_2^4}\biggr]_\epsilon
-\frac{13}{64}\left(\frac{Z\vec{r}_1}{r_1^3}-\frac{Z\vec{r}_2}{r_2^3}\right)\cdot\frac{\vec{r}}{r^3}
+\frac{1}{4}\left(\frac{Z\vec{r}_1}{r_1^3}-\frac{Z\vec{r}_2}{r_2^3}\right)\cdot\frac{\vec{r}}{r^2}-\frac{1}{4}\biggl[\frac{1}{r^3}\biggr]_\epsilon\nonumber\\
&&+\,\frac{23}{32}\biggl[\frac{1}{r^4}\biggr]_\epsilon+\frac{7}{64}\left[p_2^2,\left[p_1^2,\biggl[\frac{1}{r}\biggr]_\epsilon\right]\right]
+\frac12 \,(E-V)^3-\frac{3}{8}\,p_1^2\,(E-V)\,p_2^2\nonumber\\
&&-\,\frac{3}{8}\pi Z\biggl[\,2\left(E+\frac{Z-1}{r_2}\right)\delta^3(r_1)
+2\left(E+\frac{Z-1}{r_1}\right)\delta^3(r_2)-p_1^2\,\delta^3(r_2)-p_2^2\,\delta^3(r_1)\,\biggr]\nonumber\\
&&+\,\biggl(1-E-\frac{Z}{r_1}-\frac{Z}{r_2}-\frac{5\,\vec{P}^2}{48}\biggr)\,\pi\,\delta^3(r)
-\frac12\,\biggl[\frac{1}{2\,r}\biggl(\delta^{ij}+\frac{r^i\,r^j}{r^2}\biggr)\biggr]_\epsilon\,\nabla^i\,\nabla^j\biggl[\frac{1}{r}\biggr]_\epsilon
\nonumber\\
&&
+\frac12\,p_1^i\,\bigl(E-V\bigr)\,\frac{1}{r}\left(\delta^{ij}+\frac{r^ir^j}{r^2}\right)p_2^j
-\frac{1}{8}\frac{Z^2\,r_1^ir_2^j}{r_1^3r_2^3}\left(\frac{r^ir^j}{r}-3\,\delta^{ij}r\right)\nonumber\\
&&-\,\frac{Z}{8}\biggl[\,\frac{r_1^i}{r_1^3}\,p_2^k\left(\delta^{jk}\frac{r^i}{r}
-\delta^{ik}\frac{r^j}{r}-\delta^{ij}\frac{r^k}{r}-\frac{r^ir^jr^k}{r^3}\right)p_2^j
+(1\leftrightarrow 2)\,\biggr]\nonumber\\
&&+\,\frac{1}{8}\,p_1^k\,p_2^l\biggl[-\frac{\delta^{il}\delta^{jk}}{r}+\frac{\delta^{ik}\delta^{jl}}{r}-\frac{\delta^{ij}\delta^{kl}}{r}-\frac{\delta^{jl}r^ir^k}{r^3}
-\frac{\delta^{ik}r^jr^l}{r^3}+3\,\frac{r^ir^jr^kr^l}{r^5}\,\biggr]p_1^i\,p_2^j\nonumber\\
&&+\,\frac{1}{4}\biggl(\vec{p}_1\,\frac{1}{r^2}\,\vec{p}_1+\vec{p}_2\,\frac{1}{r^2}\,\vec{p}_2\biggr)
-\frac{1}{64}\,P^iP^j\frac{3\,r^i\,r^j-\delta^{ij}\,r^2}{r^5}
+H_{10}+H_{11}+H_{12}\biggr\rangle\nonumber\\
&&+\,\biggl\langle\frac{3}{2}\,\delta_M E\,(E-V)^2-\frac{3}{4}\,\vec{P}\,(E-V)^2\,\vec{P}-\frac{3}{8}\,\delta_M E\,p_1^2\,p_2^2+\frac{3}{16}\,P^2p_1^2p_2^2\nonumber\\
&&-\,\frac{3}{4}\,\biggl(\delta_M E+3\,E+\frac{3\,(Z-1)}{r_2}-\vec{p}_1\cdot\vec{p}_2\biggr)\pi Z\,\delta^3(r_1)+(1\leftrightarrow2)\nonumber\\
&&+\,\frac{1}{2}\,\delta_M E\,p_1^i\,\frac{1}{r}\left(\delta^{ij}+\frac{r^ir^j}{r^2}\right)p_2^j
-\frac{1}{4}\,\vec{P}^2\,p_1^i\,\frac{1}{r}\left(\delta^{ij}+\frac{r^ir^j}{r^2}\right)\,p_2^j
+\frac{13}{32}\biggl[\frac{Z^2}{r_1^4}+\frac{Z^2}{r_2^4}\biggr]_\epsilon\nonumber\\
&&+\,\frac{13}{16}\,\frac{Z^2\,\vec{r}_1\cdot\vec{r}_2}{r_1^3r_2^3}-\pi\,\delta^3(r)\biggl(\delta_ME-\frac{\vec{P}^2}{2}\biggr)
\biggr\rangle+\langle\delta_M H^{(6)}\rangle\,,
\end{eqnarray}
where
\begin{align}\label{47}
\langle\delta_M H^{(6)}\rangle=&
\biggl\langle \frac{Z}{2}\left[p_1^i\,(E-V)\left(\frac{\delta^{ij}}{r_1}+\frac{r_1^ir_1^j}{r_1^3}\right)
+p_2^i\,(E-V)\left(\frac{\delta^{ij}}{r_2}+\frac{r_2^ir_2^j}{r_2^3}\right)\right] P^j
\nonumber\\
&-\,\frac{Z}{4}\biggl[\,p_1^i\,p_2^k\left(\frac{\delta^{ij}}{r_1}+\frac{r_1^ir_1^j}{r_1^3}\right)p_2^k\, P^j
+p_2^i\,p_1^k\left(\frac{\delta^{ij}}{r_2}+\frac{r_2^ir_2^j}{r_2^3}\right)p_1^k\,P^j\,\biggr]
-\frac{Z^2}{2}\,\frac{\vec{r}_1\cdot\vec{r}_2}{r_1^3r_2^3}\nonumber\\
&+\,\frac{Z}{4}\biggl[\,p_2^i\left(\frac{\delta^{ij}}{r}+\frac{r^ir^j}{r^3}\right)\left(\frac{\delta^{jk}}{r_1}+\frac{r_1^jr_1^k}{r_1^3}\right)+
p_1^i\left(\frac{\delta^{ij}}{r}+\frac{r^ir^j}{r^3}\right)\left(\frac{\delta^{jk}}{r_2}+\frac{r_2^jr_2^k}{r_2^3}\right)\biggr]P^k\nonumber\\
&+\,\frac{Z^2}{4}\biggl[\,\vec{p}_1\,\frac{1}{r_1^2}\,\vec{p}_1
+\vec{p}_2\,\frac{1}{r_2^2}\,\vec{p}_2
+p_1^i\left(\frac{\delta^{ij}}{r_1}+\frac{r_1^ir_1^j}{r_1^3}\right)\left(\frac{\delta^{jk}}{r_2}+\frac{r_2^jr_2^k}{r_2^3}\right)p_2^k\biggr]\nonumber\\
&+\,\frac{Z^3\,\vec{r}_1\cdot\vec{r}_2}{4r_1^3r_2^2}+\frac{Z^3\,\vec{r}_1\cdot\vec{r}_2}{4r_1^2r_2^3}
+\frac{Z^2}{8}\left(\frac{r_1^i}{r_1^3}+\frac{r_2^i}{r_2^3}\right)\biggl(\frac{r_1^ir_1^j-3\,\delta^{ij}\,r_1^2}{r_1}
-\frac{r_2^ir_2^j-3\,\delta^{ij}\,r_2^2}{r_2}\biggr)\frac{r^j}{r^3}\nonumber\\
&+\,\frac{Z^2}{8}\biggl[\,p_2^k\,\frac{r_1^i}{r_1^3}\left(-\delta^{ik}\frac{r_2^j}{r_2}+\delta^{jk}\frac{r_2^i}{r_2}-\delta^{ij}\frac{r_2^k}{r_2}-\frac{r_2^ir_2^jr_2^k}{r_2^3}\right)p_2^j
+(1\leftrightarrow 2)\,\biggr]\\
&+\,\frac{1}{4}\biggl[\frac{Z^3}{r_1^3}+\frac{Z^3}{r_2^3}\biggr]_\epsilon-\frac{1}{8}\biggl[\frac{Z^2}{r_1^4}+\frac{Z^2}{r_2^4}\biggr]_\epsilon
-\frac{3\,Z^3}{2}[\pi\,\delta^3(r_1)+\pi\,\delta^3(r_2)]
+\delta_M H_8+ \delta_M H_9\biggr\rangle\,,\nonumber
\end{align}
and where $H_8$ and $H_9$ are presented in Eqs. (\ref{29}) and (\ref{30}) respectively.

\section{Total recoil correction}
The final results are split into five parts: (i) the second-order and third-order matrix elements
containing $H_R$, (iii) the second-order and third-order matrix elements containing $H_C$,
(v) the first-order matrix elements between the reference state and the perturbed
wave function, and (vi) the remaining first-order terms with the exception of (vii) pure recoil, the
radiative recoil and the recoil corrections to one-loop and two-loops radiative corrections.
The final formula for singlet states of helium is then
\begin{equation}\label{48}
\delta_M E^{(6)} = E_\textrm{i} + E_\textrm{iii} + E_\textrm{v} + E_\textrm{vi}
+ E_\textrm{vii}\,,
\end{equation}
where
\begin{eqnarray}\label{Ei}
E_\textrm{i} &=&  \left\langle H_R\,\frac{1}{(E-H)'}\,\biggl(\frac{ \vec{P}^2}{2}-\delta_M E\biggr)\,\frac{1}{(E-H)'}\,H_R\right\rangle \\
&&+\,2\left\langle H_R\,\frac{1}{(E-H)'}\,[H_R-\langle H_R\rangle]\,\frac{1}{(E-H)'}\,\frac{ \vec{P}^2}{2}\right\rangle
+2\left\langle \delta_M H_R\frac{1}{(E-H)'}H_R\right\rangle,\nonumber\\
E_\textrm{iii} &=&  \left\langle H_C\,\frac{1}{(E-H)}\,\biggl(\frac{ \vec{P}^2}{2}-\delta_M E\biggr)\,\frac{1}{(E-H)}\,H_C\right\rangle\nonumber \\
&&+\,2\left\langle H_C\,\frac{1}{(E-H)}\,H_C\frac{1}{(E-H)}\,\frac{ \vec{P}^2}{2}\right\rangle
+2\left\langle \delta_M H_C\,\frac{1}{(E-H)}\, H_C\right\rangle,
\end{eqnarray}
and where $H_R$ is defined in Eq. (\ref{37}), $\delta_M H_{R}$ in Eq. (\ref{38}), and
$H_C$ and  $\delta_M H_C$ in Eq. (\ref{17}). The terms $E_\textrm{ii}$ and $E_\textrm{iv}$ vanish for singlets.
The first-order terms $\delta_M A_2$ and $\delta_M B$  become the sum of $E_\textrm{v}$, $E_\textrm{vi}$
and $E_\textrm{vii}$. In order to explicitly cancel out $1/\epsilon$ terms
and simplify the final result we perform the following further transformations
\begin{eqnarray}\label{51}
\left[p_2^2,\left[p_1^2,\biggl[\frac{1}{r}\biggr]_\epsilon\right]\right]
 &=&\left(\frac{Z\vec{r}_1}{r_1^3}-\frac{Z\vec{r}_2}{r_2^3}\right)\cdot\frac{\vec{r}}{r^3}-2\biggl[\frac{1}{r^4}\biggr]_\epsilon
+P^iP^j\,\frac{3r^ir^j-\delta^{ij}r^2}{r^5}\nonumber\\
&&-\frac43\,\pi\,\delta^d(r)\,P^2\,,\\
\biggl[\frac{1}{r^4}\biggr]_\epsilon
&=&\biggl[\frac{1}{r^3}\biggr]_\epsilon+\frac{1}{2}\left(\vec{p}_1\,\frac{1}{r^2}\,\vec{p}_1
+\vec{p}_2\,\frac{1}{r^2}\,\vec{p}_2\right)-\left(E+\frac{Z}{r_1}+\frac{Z}{r_2}\right)\frac{1}{r^2}\nonumber\\
&&-\,\frac{m}{M}\,\biggl(\delta_M E-\frac{\vec{P}^2}{2}\biggr)\frac{1}{r^2}\,,\label{52}\\
\biggl[\frac{Z^2}{r_1^4}\biggr]_\epsilon\label{53}
&=&\vec{p}_1\,\frac{Z^2}{r_1^2}\,\vec{p}_1-2\left(E+\frac{Z}{r_2}-\frac{1}{r}\right)\frac{Z^2}{r_1^2}+p_2^2\,\frac{Z^2}{r_1^2}-2\biggl[\frac{Z^3}{r_1^3}\biggr]_\epsilon\nonumber\\
&&-\,2\,\frac{m}{M}\,\biggl(\delta_M E-\frac{\vec{P}^2}{2}\biggr)\frac{Z^2}{r_1^2}\,,\\
p_1^i\left(\frac{\delta^{ij}}{r}+\frac{r^ir^j}{r^3}\right)p_2^j&=&-\,2\,H^{(4)}_M-(E-V)^2+\frac{1}{2}\,p_1^2\,p_2^2
+Z\pi\bigl[\delta^3(r_1)+\delta^3(r_2)\bigr]+2\,\pi\,\delta^3(r)\nonumber\\
&&-\,2\,\frac{m}{M}\,\biggl[\bigl(E-V\bigr)\,\biggl(\delta_M E-\frac{\vec{P}^2}{2}\biggr)-\delta_M H^{(4)}\biggr]\,,\label{54}\\
\vec{p}_1\cdot\vec{p}_2\,\biggl[\frac{1}{r}\biggr]_\epsilon\,\vec{p}_1\cdot\vec{p}_2
&=& p_1^2\,\biggl[\frac{1}{r}\biggr]_\epsilon\,p_2^2-\vec{p}_1\times\vec{p}_2\,\frac{1}{r}\,\vec{p}_1\times\vec{p}_2-2\,\pi\,\delta^d(r)\,P^2\,.\label{55}
\end{eqnarray}
The final result for $E_\textrm{v}$ and $E_\textrm{vi}$ in terms of $Q_i$ operators
defined in Tables \ref{oprsQ} - \ref{oprsdQ} is
\begin{eqnarray}\label{Ev}
E_\textrm{v}&=&
-\,\frac{E}{8}\,Z\,\delta_M\langle Q_1\rangle + \frac18\,\delta_M\langle Q_2\rangle + \frac{1}{8}\,Z\,(1-2\,Z)\,\delta_M\langle Q_3\rangle
+\frac{3}{16}\,Z\,\delta_M\langle Q_4\rangle - \frac{Z}{4}\,\delta_M\langle Q_5\rangle\nonumber\\
&&+\,\frac{1}{24}\,\delta_M\langle Q_6\rangle+\frac{E^2+2\,E^{(4)}}{4}\,\delta_M\langle Q_7\rangle -\frac{E}{2}\,\delta_M\langle Q_8\rangle
+\frac{1}{4}\,\delta_M\langle Q_9\rangle+\frac{E}{2}\,Z^2\,\delta_M\langle Q_{11}\rangle\nonumber\\
&&+\,E\,Z^2\,\delta_M\langle Q_{12}\rangle -  E\,Z\,\delta_M\langle Q_{13}\rangle -  Z^2\,\delta_M\langle Q_{14}\rangle
+  Z^3\,\delta_M\langle Q_{15}\rangle - \frac{Z^2}{2}\,\delta_M\langle Q_{16}\rangle\nonumber\\
&&-\,\frac{Z}{2}\,\delta_M\langle Q_{17}\rangle + \frac{Z}{16}\,\delta_M\langle Q_{18}\rangle
+ \frac{Z}{2}\,\delta_M\langle Q_{19}\rangle- \frac{Z^2}{8}\,\delta_M\langle Q_{20}\rangle + \frac{Z^2}{4}\,\delta_M\langle Q_{21}\rangle \nonumber\\
&& +\,\frac{Z^2}{4}\,\delta_M\langle Q_{22}\rangle + \delta_M\langle Q_{23}\rangle + \frac{Z}{2}\,\delta_M\langle Q_{24}\rangle
- \frac{1}{32}\,\delta_M\langle Q_{25}\rangle - \frac{Z}{4}\,\delta_M\langle Q_{26}\rangle \nonumber\\
&&-\,\frac{E}{8}\,\delta_M\langle Q_{27}\rangle
- \frac{Z}{2}\,\delta_M\langle Q_{28}\rangle +\frac{1}{4}\,\delta_M\langle Q_{29}\rangle + \frac{1}{8}\,\delta_M\langle Q_{30}\rangle+\delta_ME_H\,,
\end{eqnarray}
where $\delta_M E_H$ is the remainder from $H_{12}$ in Eq. (\ref{h12}) after cancellation of $1/\epsilon$ singularities,
\begin{eqnarray}\label{58}
\delta_M E_H=\biggl(-\,4\ln\alpha-\frac{39\,\zeta(3)}{\pi^2}+\frac{32}{\pi^2}-6\ln(2)+\frac{7}{3}\biggr)\,\frac{\delta_M \langle Q_2\rangle}{16}\,.
\end{eqnarray}
and
\begin{eqnarray}\label{Evi}
E_\textrm{vi}&=&\biggl\langle-\,\frac32\,E^3 -3\,E E^{(4)} - 2\,E^2\,\delta_M E-\frac{3\,E+\delta_M E+4\,Z^2}{8}\,Z\,Q_1
-\frac{Z\,(8\,Z-3)}{8}\,Q_3\nonumber\\
&&-\,\frac34\,Z\, Q_5+\frac18\, Q_6
+\frac{3\,E^2+2\,E\,\delta_M E+6\,E^{(4)}+2\,\delta_M E^{(4)}}{4}\,Q_7-\frac12\,\delta_M E\,Q_8\nonumber\\
&&+\,\frac{2\,E+\delta_M E}{2}\,Z^2\,Q_{11}
+(3\,E+\delta_M E)\,(Z^2\,Q_{12}-Z\,Q_{13})
-3\,Z^2\,Q_{14}+\frac52\,Z^3\,Q_{15}\nonumber\\
&&-\,Z^2\,Q_{16}+\frac32\,Z\,Q_{17}+Z^2\,Q_{21}+\frac32\,Z^2\,Q_{22}+\frac32\,Z\,Q_{24}
-\frac{1}{8}\,\delta_M E\,Q_{27}-\frac34\,Z\,Q_{28}\nonumber\\
&&+\,\frac38\,Z\, Q_{31}+\frac{Z^2}{8} \,Q_{32}-\,\frac32\,E\,Z\,Q_{34}+\frac{E}{2}\,Q_{35}
-\frac34\,Z^2\,Q_{36}-Z^2\,Q_{37}+\frac32\,Z\,Q_{38}\nonumber\\
&&+\,\frac{3}{16}\,Q_{40}-\frac14\,Q_{41}+\frac{Z^2}{2}\,Q_{42}+\frac{Z^2}{2}\,Q_{43}  - \frac{Z}{2}\,Q_{44}
+\frac{Z}{2}\,Q_{45} + \frac{Z^2}{4}\,Q_{46}+\frac{Z^3}{2}\,Q_{47} \nonumber\\
&& +\, \frac{Z^2}{4}\,Q_{48}-\frac{Z^2}{4}\,Q_{49} + \frac{Z^2}{4}\,Q_{50}\biggr\rangle\,.
\end{eqnarray}
Finally,
\begin{equation}\label{Evii}
E_\textrm{vii}=\langle \delta_M H_8+ \delta_M H_9\,\rangle+\delta_M\langle \,H_{10}+H_{11}\,\rangle.
\end{equation}

\section{Numerical results}
The numerical calculations of the nonrelativistic energy and wave function were performed in the
explicitly correlated exponential basis with nonlinear parameters generated randomly within
variationally optimized intervals, a method described in the literature by Korobov \cite{Korobov}.
The method is very efficient and allows getting accuracy for energies as high as 16 digits with a
basis as small as 1500 functions. The evaluation of second-order matrix elements is more
complicated and requires large values of nonlinear parameters for obtaining accurate results. In
order to avoid numerical problems related to linear dependence in the basis set, all the
calculations are performed in octuple precision arithmetics.

Table \ref{oprsQ} presents our results for the expectation values of operators $Q_{i=1,\ldots ,30}$
which appear in the evaluation of the nonrecoil $\alpha^6\,m$ corrections for singlet states of
helium. Table \ref{oprsQ2} presents results for the expectation values of additional operators
$Q_{i=31,\ldots ,50}$ which appear in the recoil correction to order $\alpha^6\,m^2/M$. Table
\ref{oprsdQ} presents results for the matrix elements of $Q_{i=1,\ldots ,30}$ perturbed by the
nuclear kinetic energy operator. These are all matrix elements that are needed to obtain energy
shifts of order $\alpha^6\,m$ and $\alpha^6\,m^2/M$. Table \ref{Es} presents the results for the
individual contributions to the recoil $\alpha^6\,m^2/M$ correction. We notice that the photon
exchange contributions $E_{\rm i}+E_{\rm iii}+E_{\rm v}+E_{\rm vi}$ tend to cancel each other and
their net effect is relatively small in comparison to $E_{\rm vii}$. Only for the $2^1P_1$ state are both
parts of the same order. Table \ref{tab:is} presents our summary of all contributions to the
isotope shift in the $2^1S - 2^3S$ transition for a point nucleus. It includes two additional
contributions. The first one is a small shift due to the nuclear polarizability. The second
contribution is due to the hyperfine mixing of  $2^1S$ and $2^3S$ levels, which is a nominally
$\alpha^6\,m^3/M^2$ correction, but is enhanced by a small energy difference between these states.

In Table \ref{tab:abs} we present the status of the theoretical prediction of the $2^1S-2^3S$
transition energy of $^4$He. All contributions listed in the table are numerically exact
\cite{yerokhin:10:helike}, except for $\alpha^7\,m$. Following Refs. \cite{yerokhin:10:helike},
this contribution is estimated based on the known hydrogenic result. Due to a strong cancellation
of the estimate between the $2^1S$ and $2^3S$ states, the uncertainty of the difference is
difficult to guess, so we assumed $50\%$ of the whole contribution. We observe a fair agreement
with the experimental value from Ref. \cite{rooij:11}. In fact, the difference with the experiment 
will be 10 times smaller, if we neglect the $\alpha^7\,m$ contribution completely, 
so we may have overestimated its magnitude.

\section{Nuclear charge radius difference}

We now turn to the determination of the nuclear charge radii difference from the isotope shift.
Table \ref{tab:is} presents theoretical results for individual contributions to the isotope shift
in the $2^1S-2^3S$ transition, for the point nucleus. The contribution of the higher-order
$\alpha^7\,m^2/M$ QED effects was estimated on the basis of the double logarithmic contribution to
the Lamb shift in hydrogen, which for helium takes the form \cite{eides}
\begin{equation}
E^{(7)} \approx -Z^3\,\alpha^7\,\ln^2\,(Z\alpha)^{-2}\,m\,\langle\delta^3(r_1) + \delta^3(r_2)\rangle_M
\end{equation}
and we ascribe a 50\% uncertainty to this estimate. The total uncertainty of the theoretical
prediction amounts to just 0.2~kHz, which is an order of magnitude smaller than the present
experimental error, see Table \ref{tab:rms:roij}.

By comparing the theoretical (point-nucleus) and experimental values of the centroid energies of a
transition in $^3$He and $^4$He, we extract the difference in the squares of the nuclear charge
radii, $\delta r^2 = r^2(^3\mbox{\rm He}) - r^2(^4\mbox{\rm He})$. The difference between the
theoretical point-nucleus result and the measured isotope shift frequency can be ascribed solely to
the finite nuclear size shift, which can be parameterized as $ E_{\rm fs} = C\,r^2$, with $C$ being
a parameter calculated numerically. Using the experimental results for the $2^1S-2^3S$ transition
energies in $^3$He and $^4$He from Ref.~\cite{rooij:11} and taking into account the experimental
hyperfine shift of the $ 2^3S_1$ state, we obtain $\delta r^2$ as described in
Table~\ref{tab:rms:roij}, with the result $\delta r^2 = 1.027\,(11)\ \mbox{\rm fm$^2$}$. It does
not agree with the  $\delta r^2$ values obtained in Ref. \cite{herecoil, heis} from the isotope
shift in the $2^3P$-$2^3S$ transition, namely $\delta r^2 = 1.069\,(3)\ \mbox{\rm fm$^2$}$
\cite{cancio:04,cancio:12:3he} and $\delta r^2  = 1.061\,(3)\; {\rm fm}^2$ \cite{shiner:95}. We
observe that the two results from the $2^3P-2^3S$ transitions are in only slight disagreement with
each other but both deviate significantly from the result obtained from the $2^1S-2^3S$ transition.

\section{Summary}
The $4\,\sigma$ discrepancy for $\delta r^2$ is very puzzling, since we cannot explain it by any
missed corrections in the theoretical predictions. All significant theoretical contributions have
been calculated and the theoretical uncertainty is orders of magnitude smaller than the deviation.
This discrepancy calls for the verification of the experimental transition frequencies (first of
all, $2^1S-2^3S$) by independent measurements. Moreover, it can be also accessed by isotope shift
measurements in muonic helium. Hopefully, this might be accomplished in the next measurement of the
Lamb shift in muonic helium at the Paul Scherrer Institute by the CREMA Collaboration
\cite{antognini:11}. This experiment will provide an independent determination of the charge radii
of helium isotopes, thus shedding light on the proton charge radius puzzle and on the discrepancy
for the helium nuclear charge radius difference.

\begin{acknowledgments}
We wish to thank Grzegorz \L ach for his interesting discussions.
K.P. and V.P. acknowledge support by the National Science Center (Poland) Grant No.
2012/04/A/ST2/00105, and V.A.Y. acknowledges support by the Ministry of Education and Science of
the Russian Federation (program for organizing and carrying out scientific investigations) and by
the RFBR (Grant No. 16-02-00538).
\end{acknowledgments}

\appendix

\section{Derivation of $\delta_M A_2$}

$A_2^M$ is split into six parts in the order that they appear in  Eq. (\ref{43})
\begin{equation}\label{A1}
A_2^M= A_{2a}^M+A_{2b}^M+A_{2c}^M+A_{2d}^M+A_{2e}^M+A_{2f}^M.
\end{equation}
The first three terms contain both recoil and nonrecoil parts while the latter three contain only recoil terms.
Individual parts are transformed as follows:
\begin{eqnarray}\label{A4}
A_{2a}^M&=&\langle Q\,(H_M-E_M)\,Q\rangle_M=\frac{1}{2}\,\langle [Q,[H_M-E_M,Q]]\rangle_M\nonumber\\
&=&\frac12 \,\langle(\nabla_1 Q)^2+(\nabla_2 Q)^2\rangle_M + \frac14 \frac{m}{M}\langle[Q,[\vec{P}^2,Q]]\rangle\nonumber\\
&=&\biggl\langle\frac{1}{32}\biggl[\frac{Z^2}{r_1^4}+\frac{Z^2}{r_2^4}\biggr]_\epsilon+\frac{(d-1)^2}{16}\biggl[\frac{1}{r^4}\biggr]_\epsilon
-\frac{Z}{8}\biggl(\frac{\vec{r}_1}{r_1^3}-\frac{\vec{r}_2}{r_2^3}\biggr)\cdot\frac{\vec{r}}{r^3}\biggr\rangle_M\nonumber\\
&&+\,\frac{m}{M}\biggl\langle\frac{1}{32}\biggl[\frac{Z^2}{r_1^4}+\frac{Z^2}{r_2^4}\biggr]_\epsilon+\frac{1}{16}\frac{Z^2\vec{r}_1\cdot\vec{r}_2}{r_1^3r_2^3}\biggr\rangle,\\
A_{2b}^M&=&2\,E^{(4)}\langle Q\rangle_M+2\,\delta_ME^{(4)}\biggl(\frac{E}{2}+\biggl\langle\frac{1}{4r}\biggr\rangle\biggr),\label{A5}\\
A_{2c}^M&=& -2\,\langle H_M^{(4)}\,Q\rangle_M=X_1+X_2+X_3+X_4,\label{A6}
\end{eqnarray}
where
\begin{eqnarray}
X_4&=&-2\,\langle\delta_M H^{(4)}\,Q\rangle\nonumber\\
&=&\sum_a\biggl\langle -\,\frac{Z}{4}\,P^i\left(\frac{\delta^{ij}}{r_a}+\frac{r_a^ir_a^j}{r_a^3}\right)
\left(\frac{Z}{r_1}+\frac{Z}{r_2}-\frac{2}{r}\right)p_a^j
-\frac{Z}{4}\biggl[\frac{1}{2\,r_a}\biggl(\delta^{ij}+\frac{r_a^ir_a^j}{r_a^2}\biggr)\biggr]_\epsilon
\biggl[p_a^i,\biggl[p_a^j,\biggl[\frac{Z}{r_a}\biggr]_\epsilon\biggr]\biggr]\biggr\rangle\nonumber\\
&=&\sum_a\biggl\langle-\,\frac{Z}{4}\,P^i\biggl(\frac{\delta^{ij}}{r_a}+\frac{r_a^ir_a^j}{r_a^3}\biggr)\biggl(\frac{Z}{r_1}+\frac{Z}{r_2}-\frac{2}{r}\biggr)p_a^j
+\frac{1}{4}\biggl[\frac{Z^2}{r_a^4}\biggr]_\epsilon + \frac{Z^3}{2}\,\pi\delta^3(r_a)\biggr\rangle.\label{A7}
\end{eqnarray}
In the above the term with the Dirac delta function was obtained by using dimensionally regularized
representation of the Coulomb potential.
Further, using the identity $\langle\,\delta^d(x)\,\frac{1}{x}\,\rangle=0$
\begin{eqnarray}
X_{3}&=&-\,\biggl\langle\bigl[Z\,\pi\delta^3(r_1)+Z\,\pi\delta^3(r_2) + 2\,\pi\,\delta^3(r)\bigr]\,Q\biggr\rangle_M\nonumber\\
&=&\biggl\langle\frac{Z\,(Z-2)\,\pi}{4}\biggl(\frac{\delta^3(r_1)}{r_2}+\frac{\delta^3(r_2)}{r_1}\biggr)
+\,\frac{Z\,\pi}{2}\biggl(\frac{\delta^3(r)}{r_1}+\frac{\delta^3(r)}{r_2}\biggr)\biggr\rangle_M,\label{A8}\\
X_{2}&=&\biggl\langle p_1^i\,\frac{1}{r}\biggl(\delta^{ij}+\frac{r^ir^j)}{r^2}\biggr)\,p_2^j\,Q\biggr\rangle_M\label{A9}\\
&=&\biggl\langle-\,\frac14\,p_1^i\biggl(\frac{Z}{r_1}+\frac{Z}{r_2}-\frac{2}{r}\biggr)\frac{1}{r}\biggl(\delta^{ij}+\frac{r^ir^j}{r^2}\biggr)\,p_2^j
+\frac{d-1}{4}\biggl[p_1^i,\biggl[p_2^j,\biggl[\frac{1}{r}\biggr]_\epsilon\biggr]\biggr]\,
\biggl[\frac{1}{2r}\biggl(\delta^{ij}+\frac{r^ir^j}{r^2}\biggr)\biggr]_\epsilon\biggr\rangle_M,\nonumber\\
X_{1}&=&\frac14\,\bigl\langle\bigl[(p_1^2+p_2^2)-2\,p_1^2p_2^2\bigr]\,Q\bigr\rangle_M\nonumber\\
&=& \frac14\,\bigl\langle(p_1^2+p_2^2)\,Q\,(p_1^2+p_2^2)+\frac12 \,[p_1^2+p_2^2,[Q,p_1^2+p_2^2]]
-2\,p_1^2\,Q\,p_2^2-[p_1^2,[p_2^2,Q]]\bigr\rangle_M\nonumber\\
&=& X_{1A}+X_{1B}+X_{1C}+X_{1D}.\label{A10}
\end{eqnarray}
Here
\begin{eqnarray}
X_{1A}&=&\langle(E-V)^2\,Q\rangle_M+2\,\frac{m}{M}\,\biggl\langle(E-V)\,Q\,\biggl(\delta_M E-\frac{\vec{P}^2}{2}\biggr)\biggr\rangle\label{A11}\\
&=&\langle(E-V)^2\,Q\rangle_M+\frac{m}{M}\,\biggl\langle2\,\delta_M E\,(E-V)\,Q-\vec{P}\,(E-V)\,Q\vec{P}-\frac12\,[\vec{P},[\vec{P},(E-V)\,Q]]\biggr\rangle\,,\nonumber\\
X_{1B}&=&-\,\frac14\,\biggl\langle\biggl[V+\frac{m}{M}\,\frac{\vec{P}^2}{2},\biggl[p_1^2+p_2^2,\,Q\biggr]\biggr]\biggr\rangle_M\label{A12}\\
&=&\biggl\langle-\,\frac18\,\biggl[\frac{Z^2}{r_1^4}+\frac{Z^2}{r_2^4}\biggr]_\epsilon+\frac38\,\biggl(\frac{Z\vec{r}_1}{r_1^3}-\frac{Z\vec{r}_2}{r_2^3}\biggr)\cdot\frac{\vec{r}}{r^3}
-\frac{(d-1)}{4}\biggl[\frac{1}{r^4}\biggr]_\epsilon\biggr\rangle_M\nonumber\\
&&+\,\frac{m}{M}\,\biggl\langle\,\frac18\,\biggl[\frac{Z^2}{r_1^4}+\frac{Z^2}{r_2^4}\biggr]_\epsilon
+\frac14\,\frac{Z^2\,\vec{r}_1\cdot\vec{r}_2}{r_1^3r_2^3}\biggr\rangle\,,\nonumber\\
X_{1C}&=&\biggl\langle\,\frac18\,p_1^2\,\biggl(\frac{Z}{r_1}+\frac{Z}{r_2}\biggr)\,p_2^2
-\frac{(d-1)}{8}\,p_1^2\,\biggl[\frac{1}{r}\biggr]_\epsilon p_2^2\biggr\rangle_M,\label{A13}\\
X_{1D}&=&\biggl\langle-\,\frac{(d-1)}{16}\,\biggl[p_1^2,\biggl[p_2^2,\biggl[\frac{1}{r}\biggr]_\epsilon\biggr]\biggr]\biggr\rangle_M.\label{A14}
\end{eqnarray}
The remaining $A$ terms are
\begin{eqnarray}
A_{2d}^M&=&\frac{m}{M}\,\bigl\langle[Q,\,[H-E,\,\delta_M Q]]\bigr\rangle\nonumber\\
&=&\frac{m}{M}\,\bigl\langle(\nabla_1 Q)(\nabla_1\delta_M Q)+(\nabla_2 Q)(\nabla_2\delta_M Q)\bigr\rangle\nonumber\\
&=&\frac{m}{M}\,\biggl\langle-\,\frac{3}{16}\biggl[\frac{Z^2}{r_1^4}+\frac{Z^2}{r_2^4}\biggr]_\epsilon
+\frac38\,\biggl(\frac{Z\vec{r}_1}{r_1^3}-\frac{Z\vec{r}_2}{r_2^3}\biggr)
\cdot\frac{\vec{r}}{r^3}\biggr\rangle\,,\label{A15}\\
A_{2e}^M&=&\frac{m}{M}\,\biggl(\,\frac32\,E^{(4)}\,\biggl\langle\frac{1}{r}\biggr\rangle-3\,E E^{(4)}\biggr)\,,\label{A16}\\
A_{2f}^M&=&-2\,\frac{m}{M}\,\langle H_A\,\delta_M Q\rangle\label{A17}
=F_1+F_2+F_3\,,
\end{eqnarray}
where
\begin{eqnarray}
F_3&=&-\,\frac{m}{M}\,\biggl\langle\frac{3\,Z^2\,\pi}{4}\biggl(\frac{\delta^3(r_1)}{r_2}+\frac{\delta^3(r_2)}{r_1}\biggr)
+\frac32 \pi\,Z\biggl(\frac{\delta^3(r)}{r_1}+\frac{\delta^3(r)}{r_2}\biggr)\biggr\rangle,\label{A18}\\
F_2&=&\frac{m}{M}\,\biggl\langle\frac34\, p_1^i\,\biggl(\frac{Z}{r_1}+\frac{Z}{r_2}\biggr)\frac{1}{r}\biggl(\delta^{ij}+\frac{r^ir^j}{r^2}\biggr)\,p_2^j\biggr\rangle,\label{A19}\\
F_1&=&\frac14\,\frac{m}{M}\,\bigl\langle\bigl[(p_1^2+p_2^2)^2-2\,p_1^2p_2^2\bigr]\,\delta Q\bigr\rangle\nonumber\\
&=&\frac14\,\frac{m}{M}\,\bigl\langle(p_1^2+p_2^2)\,\delta Q\,(p_1^2+p_2^2)+\frac12 \,[p_1^2+p_2^2,[p_1^2+p_2^2,\delta Q]]
-2\,p_1^2\,\delta Q\,p_2^2\bigr\rangle\nonumber\\
&=&F_{1A}+F_{1B}+F_{1C}\,,\label{A20}
\end{eqnarray}
and where
\begin{eqnarray}
F_{1A}&=&\frac{m}{M}\,\biggl\langle\frac34\,(E-V)^2\biggl[\frac{Z}{r_1}+\frac{Z}{r_2}\biggr]_\epsilon\biggr\rangle\,,\label{A21}\\
F_{1B}&=&\frac{m}{M}\,\biggl\langle\frac38\,\biggl[\frac{Z^2}{r_1^4}+\frac{Z^2}{r_2^4}\biggr]_\epsilon
-\frac38\,\biggl(\frac{Z\vec{r}_1}{r_1^3}-\frac{Z\vec{r}_2}{r_2^3}\biggr)
\cdot\frac{\vec{r}}{r^3}\biggr\rangle\,,\label{A22}\\
F_{1C}&=&-\,\frac{m}{M}\,\biggl\langle\frac38\, p_1^2\,\biggl(\frac{Z}{r_1}+\frac{Z}{r_2}\biggr)\,p_2^2\biggr\rangle\,.\label{A23}
\end{eqnarray}
Taking now only the recoil part of terms $A_{2a}^M\ldots A_{2f}^M$ we obtain the following results:
\begin{eqnarray}\label{A24}
\delta_M A_{2a}&=&\delta_M\,\biggl\langle\frac{1}{32}\biggl[\frac{Z^2}{r_1^4}+\frac{Z^2}{r_2^4}\biggr]_\epsilon+\frac{(d-1)^2}{16}\biggl[\frac{1}{r^4}\biggr]_\epsilon
-\frac{Z}{8}\biggl(\frac{\vec{r}_1}{r_1^3}-\frac{\vec{r}_2}{r_2^3}\biggr)\cdot\frac{\vec{r}}{r^3}\biggr\rangle\\
&&+\,\biggl\langle\frac{1}{32}\biggl(\frac{Z^2}{r_1^4}+\frac{Z^2}{r_2^4}\biggr)+\frac{1}{16}\frac{Z^2\,\vec{r}_1\cdot\vec{r}_2}{r_1^3r_2^3}\biggr\rangle,\nonumber\\
\delta_M A_{2b}&=&2\,E^{(4)}\delta_M\,\langle Q\rangle+2\,\delta_ME^{(4)}\biggl(\frac{E}{2}+\biggl\langle\frac{1}{4r}\biggr\rangle\biggr),\label{A25}\\
\delta_M A_{2c}&=& \delta_M\,\biggl\langle\frac{Z(Z-2)\,\pi}{4}\biggl(\frac{\delta^3(r_1)}{r_2}+\frac{\delta^3(r_2)}{r_1}\biggr)
+\frac{Z\,\pi}{2}\biggl(\frac{\delta^3(r)}{r_1}+\frac{\delta^3(r)}{r_2}\biggr)\nonumber\\
&&-\,\frac14\,p_1^i\biggl(\frac{Z}{r_1}+\frac{Z}{r_2}-\frac{2}{r}\biggr)\frac{1}{r}\biggl(\delta^{ij}+\frac{r^ir^j}{r^2}\biggr)\,p_2^j\label{A26}\nonumber\\
&&+\,\frac{(d-1)}{4}\biggl[p_1^i,\biggl[p_2^j,\biggl[\frac{1}{r}\biggr]_\epsilon\biggr]\biggr]
\,\biggl[\frac{1}{2r}\biggl(\delta^{ij}+\frac{r^ir^j}{r^2}\biggr)\biggr]_\epsilon
+(E-V)^2\,Q -\frac18 \biggl[\frac{Z^2}{r_1^4}+\frac{Z^2}{r_2^4}\biggr]_\epsilon\nonumber\\
&&+\,\frac38 \biggl(\frac{Z\vec{r}_1}{r_1^3}-\frac{Z\vec{r}_2}{r_2^3}\biggr)\cdot\frac{\vec{r}}{r^3}-\frac{(d-1)}{4}\biggl[\frac{1}{r^4}\biggr]_\epsilon
+\frac18\,p_1^2\,\biggl(\frac{Z}{r_1}+\frac{Z}{r_2}\biggr)\,p_2^2\nonumber\\
&&-\,\frac{(d-1)}{8}\,p_1^2\,\biggl[\frac{1}{r}\biggr]_\epsilon\,p_2^2
-\frac{(d-1)}{16}\biggl[p_1^2,\biggl[p_2^2,\biggl[\frac{1}{r}\biggr]_\epsilon\biggr]\biggr]\biggr\rangle\nonumber\\
&&+\,\biggl\langle -\,\frac{Z}{4} \sum_aP^i\left(\frac{\delta^{ij}}{r_a}+\frac{r_a^ir_a^j}{r_a^3}\right)
\biggl(\frac{Z}{r_1}+\frac{Z}{r_2}-\frac{2}{r}\biggr)\,p_a^j
+\frac{3}{8}\biggl[\frac{Z^2}{r_1^4}+\frac{Z^2}{r_2^4}\biggr]_\epsilon\nonumber\\
&& + \,\frac{Z^3}{2}\bigl(\pi\delta^3(r_1)+\pi\delta^3(r_2)\bigr)
+2\,\delta_M E\,(E-V)\,Q-\vec{P}\,(E-V)\,Q\vec{P}\nonumber\\
&&-\,\frac12 \,[\vec{P},[\vec{P},(E-V)\,Q]]+\frac14\frac{Z^2\,\vec{r}_1\cdot\vec{r}_2}{r_1^3r_2^3}\biggr\rangle\,,\\
\delta_M A_{2d}&=&\biggl\langle-\,\frac{3}{16}\biggl[\frac{Z^2}{r_1^4}+\frac{Z^2}{r_2^4}\biggr]_\epsilon
+\frac38\biggl(\frac{Z\vec{r}_1}{r_1^3}-\frac{Z\vec{r}_2}{r_2^3}\biggr)
\cdot\frac{\vec{r}}{r^3}\biggr\rangle,\label{A27}\\
\delta_M A_{2e}&=&\frac32\,E^{(4)}\biggl\langle\frac{1}{r}\biggr\rangle-3\,E E^{(4)},\label{A28}\\
\delta_M A_{2f}&=&\biggl\langle-\,\frac{3\,Z^2\pi}{4}\biggl(\frac{\delta^3(r_1)}{r_2}+\frac{\delta^3(r_2)}{r_1}\biggr)
+\frac34\, p_1^i\,\biggl(\frac{Z}{r_1}+\frac{Z}{r_2}\biggr)\frac{1}{r}\biggl(\delta^{ij}+\frac{r^ir^j}{r^2}\biggr)\,p_2^j\nonumber\\
&&+\,\frac34\,(E-V)^2\biggl[\frac{Z}{r_1}+\frac{Z}{r_2}\biggr]_\epsilon
+\frac38\biggl[\frac{Z^2}{r_1^4}+\frac{Z^2}{r_2^4}\biggr]_\epsilon-\frac38\biggl(\frac{Z\vec{r}_1}{r_1^3}-\frac{Z\vec{r}_2}{r_2^3}\biggr)
\cdot\frac{\vec{r}}{r^3}\nonumber\\
&&-\,\frac38\, p_1^2\,\biggl(\frac{Z}{r_1}+\frac{Z}{r_2}\biggr)\,p_2^2
-\frac32 \pi\,Z\biggl(\frac{\delta^3(r)}{r_1}+\frac{\delta^3(r)}{r_2}\biggr)\biggr\rangle\,. \label{A29}
\end{eqnarray}
Summing all of the recoil parts $\delta_M A_{2a}\ldots \delta_M A_{2f}$ and using the identity
\begin{eqnarray}\label{A30}
[\vec{P},[\vec{P},(E-V)\,Q]]&=&\frac12\,\biggl[\frac{Z^2}{r_1^4}+\frac{Z^2}{r_2^4}\biggr]_\epsilon+\frac{Z^2\,\vec{r}_1\cdot\vec{r}_2}{r_1^3r_2^3}
-\biggl(E+\frac{2\,Z-3}{r_2}\biggr)\pi\,Z\,\delta^3(r_1)\nonumber\\
&&-\,\biggl(E+\frac{2\,Z-3}{r_1}\biggr)\pi\,Z\,\delta^3(r_2)
\end{eqnarray}
we get the final result in Eq. (\ref{44}).

\section{Derivation of $\delta_M B$}
In the following we perform only derivation of terms $B_1^M\ldots\,B_7^M$,
defined as
\begin{equation}\label{B1}
B_i^M =  \langle H_i^M\rangle_M
\end{equation}
and the evaluation of the remaining terms is trivial since they contain only Dirac delta-like contributions.
The expectation value of the kinetic energy term
\begin{equation}\label{B2}
H_1^M=\frac{1}{16}\,\bigl(p_1^6+p_2^6\bigr)
\end{equation}
is
\begin{eqnarray}\label{B3}
B_1^M&=&\frac{1}{16}\,\bigl\langle(p_1^2+p_2^2)^3-3\,p_1^2p_2^2\,(p_1^2+p_2^2)\bigr\rangle_M\nonumber\\
&=&\biggl\langle\frac{1}{8}\,\biggl[V+\frac{m}{M}\frac{\vec{P}^2}{2},\biggl[p_1^2+p_2^2,V\biggr]\biggr]
+\frac{1}{2}\,\biggl(E-V+\frac{m}{M}\biggl(\delta_M E-\frac{\vec{P}^2}{2}\biggr)\biggr)^3\nonumber\\
&&-\,\frac{3}{8}\,p_1^2\,p_2^2\,\biggl(E-V+\frac{m}{M}\biggl(\delta_M E-\frac{\vec{P}^2}{2}\biggr)\biggr)\biggr\rangle_M \nonumber\\
&=& \biggl\langle\frac14\,\bigl[(\nabla_1 V)^2+(\nabla_2 V)^2\bigr]+\frac12\,(E-V)^3-\frac38\,p_1^2\,(E-V)\,p_2^2\nonumber\\
&&+\,\frac{3}{16}\,[p_1^2,[p_2^2,V]]\biggr\rangle_M+\frac{m}{M}\biggl\langle\frac32\,(E-V)^2\biggl(\delta_M E-\frac{\vec{P}^2}{2}\biggr)
-\frac38\,p_1^2p_2^2\biggl(\delta_M E-\frac{\vec{P}^2}{2}\biggr)\nonumber\\
&&-\,\frac12\biggl[\frac{Z^2}{r_1^4}+\frac{Z^2}{r_2^4}\biggr]_\epsilon-\frac{Z^2\,\vec{r}_1\cdot\vec{r}_2}{r_1^3r_2^3}\biggr\rangle\,.
\end{eqnarray}
Recoil correction $\delta_M B_1$ is then
\begin{eqnarray}\label{B4}
\delta_M B_1&=&\delta_M\,\biggl\langle\frac14\biggl[\frac{Z^2}{r_1^4}+\frac{Z^2}{r_2^4}\biggr]_\epsilon
-\frac12 \biggl(\frac{Z\vec{r}_1}{r_1^3}-\frac{Z\vec{r}_2}{r_2^3}\biggr)\cdot\frac{\vec{r}}{r^3}
+\frac{1}{2}\biggl[\frac{1}{r^4}\biggr]_\epsilon+\frac12\,(E-V)^3\nonumber\\
&&+\,\frac{3}{16}\,\biggl[p_1^2,\biggl[p_2^2,\biggl[\frac{1}{r}\biggr]_\epsilon\biggr]\biggr]
-\frac{3}{8}\,p_1^2\,(E-V)\,p_2^2\biggr\rangle\nonumber\\
&&+\,\biggl\langle\frac32\,\delta_M E\,(E-V)^2-\frac34\,\vec{P}\,(E-V)^2\,\vec{P}+\frac14\biggl[\frac{Z^2}{r_1^4}+\frac{Z^2}{r_2^4}\biggr]_\epsilon
+\frac12\frac{Z^2\,\vec{r}_1\cdot\vec{r}_2}{r_1^3r_2^3}\nonumber\\
&&-\,3\,\biggl(E+\frac{Z-1}{r_2}\biggr)\,\pi\,Z\,\delta^3(r_1)+(1\leftrightarrow2)-\frac38\,p_1^2p_2^2\,\biggl(\delta_M E-\frac{\vec{P}^2}{2}\biggr)\biggr\rangle.
\end{eqnarray}
Here we used the identity
\begin{eqnarray}\label{B5}
[\vec{P},[\vec{P},(E-V)^2]]&=&-\,2\,\biggl[\frac{Z^2}{r_1^4}+\frac{Z^2}{r_2^4}\biggr]_\epsilon-4\,\frac{Z^2\,\vec{r}_1\cdot\vec{r}_2}{r_1^3r_2^3}\nonumber\\
&&+\,2\,(E-V)\,\bigl[4\pi\,Z\delta^3(r_1)+4\pi\,Z\delta^3(r_2)\bigr].
\end{eqnarray}
The operator $H^M_2$ is
\begin{equation}\label{B6}
H_2^M=\sum_{a=1,2}\frac{(\nabla_a V)^2}{8}+\frac{5}{128}\,\bigl[p_a^2,\bigl[p_a^2,V\bigr]\bigr]-\frac{3}{64}\,\bigl\{p_a^2,\nabla_a^2 V\bigr\}.
\end{equation}
For the sake of simplicity we split its expectation value into three parts,
\begin{eqnarray}
B_2^M&&=\biggl\langle\,\frac{1}{8}\,\bigl[(\nabla_1 V)^2+(\nabla_2 V)^2\bigr]
+\frac{5}{128}\,\bigl(\bigl[p_1^2,\bigl[p_1^2,V\bigr]\bigr]+\bigl[p_2^2,\bigl[p_2^2,V\bigr]\bigr]\bigr)
-\frac{3}{32}\,\bigl(p_1^2\,\nabla_1^2V+p_2^2\,\nabla_2^2V\bigr)\biggr\rangle_M\nonumber\\
&&=B_{2a}^M+B_{2b}^M+B_{2c}^M.\label{B7}
\end{eqnarray}
The term
\begin{equation}
B_{2a}^M=\frac18\langle(\nabla_1V)^2+(\nabla_2V)^2\rangle_M
\end{equation}
needs no further reduction.
The remaining terms could be simplified to
\begin{eqnarray}
B_{2b}^M&=&\frac{5}{128}\,\bigl\langle \bigl[p_1^2+p_2^2,\bigl[p_1^2,V\bigr]\bigr]
+\bigl[p_1^2+p_2^2,\bigl[p_2^2,V\bigr]\bigr]-2\,\bigl[p_1^2,\bigl[p_2^2,V\bigr]\bigr]\bigr\rangle_M\nonumber\\\label{B8}
&=&-\,\frac{5}{64}\,\biggl\langle\biggl[V+\frac{m}{M}\frac{\vec{P}^2}{2},\biggl[p_1^2+p_2^2,V\biggr]\biggr]+\bigl[p_1^2,\bigl[p_2^2,V\bigr]\bigr]\biggr\rangle_M,\\\label{B9}
B_{2c}^M&=&-\,\frac{3}{32}\,\bigl\langle \bigl(p_1^2+p_2^2\bigr)\,\nabla_1^2 V+\bigl(p_1^2+p_2^2\bigr)\,\nabla_2^2 V-p_2^2\,\nabla_1^2 V-p_1^2\,\nabla_2^2 V\bigr\rangle_M\\
&=&-\,\frac{3}{8}\,\pi\, \biggl\langle 2\,\biggl[E-V+\frac{m}{M}\,\biggl(\delta_M E-\frac{\vec{P}^2}{2}\biggr)\biggr]\bigl(Z\,\delta^3(r_1)+Z\,\delta^3(r_2)-\delta^3(r)\bigr)\nonumber\\
&&-\,p_1^2\,Z\,\delta^3(r_2)-p_2^2\,Z\,\delta^3(r_1)\biggr\rangle_M\nonumber.
\end{eqnarray}
Taking now only the recoil parts of individual terms we get
\begin{eqnarray}\label{B10}
\delta_M B_{2a}&=&\frac{1}{8}\,\delta_M\,\biggl\langle(\nabla_1 V)^2+(\nabla_2 V)^2\biggr\rangle,\\ \label{B11}
\delta_M B_{2b}&=&-\,\frac{5}{32}\,\delta_M\,\biggl\langle(\nabla_1 V)^2+(\nabla_2 V)^2+\frac{1}{2}\bigl[p_1^2,\bigl[p_2^2,V\bigr]\bigr]
\biggr\rangle
+\frac{5}{64}\,\bigl\langle\bigl[V,\bigl[\vec{P}^2,V\bigr]\bigr]\bigr\rangle,\\
\delta_M B_{2c}&=&-\,\frac{3}{8}\,\pi \delta_M\,\biggl\langle 2\left(E+\frac{Z-1}{r_2}\right)Z\,\delta^3(r_1)+2\left(E+\frac{Z-1}{r_1}\right)Z\,\delta^3(r_2)\nonumber\\
&&-\,2\left(E+\frac{Z}{r_1}+\frac{Z}{r_2}\right)\delta^3(r)
-p_1^2\,Z\,\delta^3(r_2)-p_2^2\,Z\,\delta^3(r_1)\biggr\rangle\nonumber\\
&&-\,\frac{3}{4}\,\pi \,\biggl\langle \biggl(\delta_M E-\frac{\vec{P}^2}{2}\biggr)\bigl(Z\,\delta^3(r_1)+Z\,\delta^3(r_2)-\delta^3(r)\bigr)\biggr\rangle\,.\label{B12}
\end{eqnarray}
The term $\delta_M B_2$ is then the sum of these three terms and is
\begin{eqnarray}\label{B13}
\delta_M B_2&=&\delta_M\,\biggl\langle
-\,\frac{1}{32}\biggl[\frac{Z^2}{r_1^4}+\frac{Z^2}{r_2^4}\biggr]_\epsilon
+\frac{1}{16}\left(\frac{Z\vec{r}_1}{r_1^3}-\frac{Z\vec{r}_2}{r_2^3}\right)\cdot\frac{\vec{r}}{r^3}-\frac{1}{16}\biggl[\frac{1}{r^4}\biggr]_\epsilon
-\frac{5}{64}\biggl[p_1^2,\biggl[p_2^2,\biggl[\frac{1}{r}\biggr]_\epsilon\biggr]\biggr]\nonumber \\
&&-\,\frac{3}{8}\,\pi \biggl[\,2\left(E+\frac{Z-1}{r_2}\right)Z\,\delta^3(r_1)+2\left(E+\frac{Z-1}{r_1}\right)Z\,\delta^3(r_2)
-\,2\left(E+\frac{Z}{r_1}+\frac{Z}{r_2}\right)\delta^3(r)\nonumber\\
&&-\,p_1^2\,Z\,\delta^3(r_2)-p_2^2\,Z\,\delta^3(r_1)\biggr]\biggr\rangle
+\biggl\langle\frac{5}{32}\biggl[\frac{Z^2}{r_1^4}+\frac{Z^2}{r_2^4}\biggr]_\epsilon+\frac{5}{16}\frac{Z^2\,\vec{r}_1\cdot\vec{r}_2}{r_1^3r_2^3}\\
&&-\,\frac{3}{4}\,\pi\biggl\{\biggl(\delta_M E-E+\frac{1-Z}{r_2}-\vec{p}_1\cdot\vec{p}_2\biggr) \,Z \,\delta^3(r_1)+(1\leftrightarrow2)
-\biggl(\delta_M E-\frac{\vec{P}^2}{2}\biggr)\delta^3(r)\biggr\}\biggr\rangle\,.\nonumber
\end{eqnarray}
The operator $H_3^M$ is
\begin{equation}\label{B14}
H_3^M=-\frac{\pi}{16}\,\nabla^2\delta^3(r)-\frac{\pi}{16}\,\delta_\perp^{ij}\,P^i\,P^j
+\frac{\pi}{4}\,\delta_\perp^{ij}\,p^i\,p^j
\end{equation}
and its expectation value is
\begin{eqnarray}\label{B15}
B_3^M&=&\biggl\langle-\frac{\pi}{8}\,\nabla^2\delta^3(r)-\frac{1}{64}\,P^i\,P^j\frac{3r^3\,r^j-\delta^{ij}\,r^2}{r^5}
-\frac{\pi}{24}\,\delta^3(r)\,\vec P^2\nonumber\\
&&-\,\frac{1}{64}\,\biggl(\frac{Z\,\vec{r}_1}{r_1^3}-\frac{Z\,\vec{r}_2}{r_2^3}\biggr)\cdot\frac{\vec{r}}{r}
+\frac{1}{32}\biggl[\frac{1}{r^4}\biggr]_\epsilon\biggr\rangle_M
\end{eqnarray}
where we used the identities
\begin{eqnarray}\label{B16}
4\,\pi\,\delta_\perp^{ij}P^i\,P^j&=&P^i\,P^j\frac{3r^3\,r^j-\delta^{ij}\,r^2}{r^5}+\frac{8\,\pi}{3}\,\delta^3(r)\,\vec P^2\,,\\
4\,\pi\,\delta_\perp^{ij}p^i\,p^j&=&-\pi\,\nabla^2\,\delta^3(r)
-\frac{1}{4}\,\biggl(\frac{Z\,\vec{r}_1}{r_1^3}-\frac{Z\,\vec{r}_2}{r_2^3}\biggr)\cdot\frac{\vec{r}}{r}
+\frac{1}{2}\biggl[\frac{1}{r^4}\biggr]_\epsilon\,.\label{B17}
\end{eqnarray}
Further, with the help of identity valid for singlet states
\begin{eqnarray}\label{B18}
\langle\,\nabla^2\,\delta^3(r)\rangle_M=-2\,\biggl\langle\,\delta^3(r)\biggl(E+\frac{Z}{r_1}+\frac{Z}{r_2}-\frac{\vec P^2}{4}+\frac{m}{M}
\biggl(\delta_M\,E-\frac{\vec P^2}{2}\biggr)\biggr)\,\biggr\rangle_M
\end{eqnarray}
we get the following recoil correction $\delta_M B_3$
\begin{eqnarray}\label{B19}
\delta_M B_3&=&\delta_M\,\biggl\langle\frac{\pi}{4}\,\,\delta^3(r)\biggl(E+\frac{Z}{r_1}+\frac{Z}{r_2}-\frac{\vec P^2}{4}\biggr)
-\frac{1}{64}\,P^i\,P^j\frac{3r^3\,r^j-\delta^{ij}\,r^2}{r^5}
-\frac{\pi}{24}\,\delta^3(r)\,\vec P^2\nonumber\\
&&-\,\frac{1}{64}\,\biggl(\frac{Z\,\vec{r}_1}{r_1^3}-\frac{Z\,\vec{r}_2}{r_2^3}\biggr)\cdot\frac{\vec{r}}{r}
+\frac{1}{32}\biggl[\frac{1}{r^4}\biggr]_\epsilon\biggr\rangle
+\biggl\langle\frac{\pi}{4}\,\,\delta^3(r)\biggl(\delta_M\,E-\frac{\vec P^2}{2}\biggr)\biggr)\,\biggr\rangle\,.
\end{eqnarray}
We split the correction due to operator $H_4^M=H_4+\frac{m}{M}\,\delta_M H_4$ into two parts: the recoil correction
to operator $H_4$, which we denote $B^M_{4a}$, and the expectation value of the recoil part $\delta_M H_4$ which we
denote $B_{4b}^M$.
The nonrecoil part of the operator $H_4^M$ is
\begin{equation}\label{B20}
H_4=\frac{1}{4}\,\bigl(p_1^2+p_2^2\bigr)\,p_1^i\frac{1}{r}\left(\delta^{ij}+\frac{r^ir^j}{r^2}\right)p_2^j
-\frac{1}{4}\,(p_1^2+p_2^2)\,4\,\pi\,\delta^3(r)\,.
\end{equation}
The expectation value of this is
\begin{eqnarray}
B_{4a}^M
&=&\frac{1}{2}\,\biggl\langle\bigl(E-V\bigr)\,p_1^i\,\frac{1}{r}\biggl(\delta^{ij}+\frac{r^ir^j}{r^2}\biggr)\,p_2^j
-\frac12\,(E-V)\,4\,\pi\,\delta^3(r)\biggr\rangle_M\nonumber\\
&&+\,\frac{m}{M}\,\biggl\langle\,\frac12\,\biggl(\delta_M E-\frac{\vec{P}^2}{2}\biggr)\,p_1^i\,\frac{1}{r}\left(\delta^{ij}+\frac{r^ir^j}{r^2}\right)p_2^j
-\frac12\,\biggl(\delta_M\,E-\frac{\vec P^2}{2}\biggr)\,4\,\pi\,\delta^3(r)\biggr\rangle\nonumber\\
&=&\biggl\langle \frac{1}{2}\,p_1^i\,\bigl(E-V\bigr)\,\frac{1}{r}\biggl(\delta^{ij}+\frac{r^ir^j}{r^2}\biggr)\,p_2^j
-\frac{1}{2}\biggl[\frac{1}{2r}\biggl(\delta^{ij}+\frac{r^ir^j}{r^2}\biggr)\biggr]_\epsilon
\nabla^i\,\nabla^j\biggl[\frac{1}{r}\biggr]_\epsilon
-\frac12\,(E-V)\,4\,\pi\,\delta^3(r)\biggr\rangle_M\nonumber\\
&&+\,\frac{m}{M}\,\biggl\langle\,\frac12\,\biggl(\delta_M E-\frac{\vec{P}^2}{2}\biggr)\,p_1^i\,\frac{1}{r}\left(\delta^{ij}+\frac{r^ir^j}{r^2}\right)p_2^j
-\frac12\,\biggl(\delta_M\,E-\frac{\vec P^2}{2}\biggr)\,4\,\pi\,\delta^3(r)\biggr\rangle\,.\label{B21}
\end{eqnarray}
The recoil correction $\delta_M B_{4a}$ is then
\begin{eqnarray}
\delta_M B_{4a}&=&\delta_M\,\biggl\langle\,\frac12 \,p_1^i\,\bigl(E-V\bigr)\,\frac{1}{r}\biggl(\delta^{ij}+\frac{r^ir^j}{r^2}\biggr)p_2^j\
-\frac12\biggl[\frac{1}{2r}\biggl(\delta^{ij}+\frac{r^ir^j}{r^2}\biggr)\biggr]_\epsilon\nabla^i\,\nabla^j\,\biggl[\frac{1}{r}\biggr]_\epsilon\nonumber\\
&&-\,\frac12\,(E-V)\,4\,\pi\,\delta^3(r)\biggr\rangle
+\,\biggl\langle\,\frac12\,\biggl(\delta_M E-\frac{\vec{P}^2}{2}\biggr)\,p_1^i
\,\frac{1}{r}\biggl(\delta^{ij}+\frac{r^ir^j}{r^2}\biggr)\,p_2^j\nonumber\\
&&-\,\frac12\,\biggl(\delta_M\,E-\frac{\vec P^2}{2}\biggr)\,4\,\pi\,\delta^3(r)\biggr\rangle.\label{B22}
\end{eqnarray}
The recoil part of $H_4^M$ is
\begin{equation}\label{B23}
\delta_M H_4=\,\frac{Z}{4}\,\biggl(p_1^2\,p_1^i\biggl(\frac{\delta^{ij}}{r_1}+\frac{r_1^ir_1^j}{r_1^3}\biggr)\,P^j
+p_2^2\,p_2^i\biggl(\frac{\delta^{ij}}{r_2}+\frac{r_2^ir_2^j}{r_2^3}\biggr)\,P^j\biggr).
\end{equation}
The expectation value of this operator can then be reduced to
\begin{eqnarray}
\delta_M B_{4b}&=&\,\frac{Z}{4}\,\biggl\langle2\,\bigl(E-V\bigr)\biggl[p_1^i\biggl(\frac{\delta^{ij}}{r_1}+\frac{r_1^ir_1^j}{r_1^3}\biggr)\,P^j
+p_2^i\biggl(\frac{\delta^{ij}}{r_2}+\frac{r_2^ir_2^j}{r_2^3}\biggr)\,P^j\,\biggr]\nonumber\\
&&-\,\biggl[p_2^2\,p_1^i\biggl(\frac{\delta^{ij}}{r_1}+\frac{r_1^ir_1^j}{r_1^3}\biggr)\,P^j
+p_1^2\,p_2^i\biggl(\frac{\delta^{ij}}{r_2}+\frac{r_2^ir_2^j}{r_2^3}\biggr)\,P^j\,\biggr]\biggr\rangle\nonumber\\
&=&\biggl\langle\,\frac{Z}{2}\left[p_1^i\,\bigl(E-V\bigr)\left(\frac{\delta^{ij}}{r_1}+\frac{r_1^ir_1^j}{r_1^3}\right) P^j
+p_2^i\,\bigl(E-V\bigr)\left(\frac{\delta^{ij}}{r_2}+\frac{r_2^ir_2^j}{r_2^3}\right) P^j\right]\nonumber\\
&&-\,\frac{1}{2}\left[\frac{Z^2}{r_1^4}+\frac{Z^2}{r_2^4}\right]_\epsilon-Z^3\left[\pi\delta^3(r_1)+\pi\delta^3(r_2)\right]\nonumber\\
&&-\,\frac{Z}{4}\biggl[\,p_1^i\,p_2^k\,\left(\frac{\delta^{ij}}{r_1}+\frac{r_1^ir_1^j}{r_1^3}\biggr)\,p_2^k\, P^j
+p_2^i\,p_1^k\,\biggl(\frac{\delta^{ij}}{r_2}+\frac{r_2^ir_2^j}{r_2^3}\right)\,p_1^k\, P^j\,\biggr]\biggr\rangle\,.\label{B24}
\end{eqnarray}
The operator $H_5^M$ is
\begin{equation}\label{B25}
H_5^M=\frac{1}{2}\left(\frac{Z\vec{r}_1}{r_1^3}-\frac{Z\vec{r}_2}{r_2^3}\right)\cdot\frac{\vec{r}}{r^3}-\frac{(d-1)}{2}\biggl[\frac{1}{r^4}\biggr]_\epsilon
-\frac{(d-1)}{32}\left(\left[p_1^2,\left[p_1^2,\biggl[\frac{1}{r}\biggr]_\epsilon\right]\right]+\left[p_2^2,\left[p_2^2,\biggl[\frac{1}{r}\biggr]_\epsilon\right]\right]\right)
\end{equation}
and its expectation value is
\begin{equation}\label{B26}
B_5^M
=\biggl\langle\frac{1}{2}\,\biggl(\frac{Z\vec{r}_1}{r_1^3}-\frac{Z\vec{r}_2}{r_2^3}\biggr)\cdot\frac{\vec{r}}{r^3}-\frac{(d-1)}{2}\biggl[\frac{1}{r^4}\biggr]_\epsilon
+\frac{(d-1)}{16}\,\biggl(\biggl[V,\biggl[p_1^2+p_2^2,\biggl[\frac{1}{r}\biggr]_\epsilon\biggr]\biggr]
+\biggl[p_1^2,\biggl[p_2^2,\biggl[\frac{1}{r}\biggr]_\epsilon\biggr]\biggr]\biggr)\biggr\rangle_M\,.
\end{equation}
The recoil correction is then
\begin{eqnarray}\label{B27}
\delta_M B_5&=&\delta_M\,\biggl\langle\frac{1}{4}\left(\frac{Z\vec{r}_1}{r_1^3}-\frac{Z\vec{r}_2}{r_2^3}\right)\cdot\frac{\vec{r}}{r^3}
-\frac{(d-1)}{4}\biggl[\frac{1}{r^4}\biggr]_\epsilon
+\frac{(d-1)}{16}\,\biggl[p_1^2,\biggl[p_2^2,\biggl[\frac{1}{r}\biggr]_\epsilon\biggr]\biggr]\biggr\rangle.
\end{eqnarray}
The operator $H_6^M$ contains the recoil part $\delta_M H_6$ and we thus
again split the calculation into two parts: the recoil correction due
to $H_6$, which we denote as $\delta_M B_{6a}$, and the expectation value of
$\delta_M H_6$, which we denote as $\delta_M B_{6b}$.
The nonrecoil part of the operator $H_6^M$ is
\begin{equation}\label{B28}
H_6=\frac{1}{8}\,p_1^i\,\frac{1}{r^2}\biggl(\delta^{ij}+3\,\frac{r^ir^j}{r^2}\biggr)\,p_1^j
+\frac{1}{8}\,p_2^i\,\frac{1}{r^2}\biggl(\delta^{ij}+3\,\frac{r^ir^j}{r^2}\biggr)\,p_2^j+\frac{(d-1)}{4}\biggl[\frac{1}{r^4}\biggr]_\epsilon
\end{equation}
and the recoil correction due to it is simply
\begin{equation}\label{B29}
\delta_M B_{6a}=\delta_M\,\biggl\langle\frac{1}{8}\,p_1^i\,\frac{1}{r^2}\biggl(\delta^{ij}+3\,\frac{r^ir^j}{r^2}\biggr)\,p_1^j
+\frac{1}{8}\,p_2^i\,\frac{1}{r^2}\biggl(\delta^{ij}+3\frac{r^ir^j}{r^2}\biggr)\,p_2^j+\frac{(d-1)}{4}\biggl[\frac{1}{r^4}\biggr]_\epsilon\biggr\rangle.
\end{equation}
The expectation value of $\delta_M H_6$ is
\begin{eqnarray}
\delta_M B_{6b}&=&\biggl\langle\frac{Z}{4}\biggl[\,p_2^i\biggl(\frac{\delta^{ij}}{r}
+\frac{r^ir^j}{r^3}\biggr)\biggl(\frac{\delta^{jk}}{r_1}+\frac{r_1^jr_1^k}{r_1^3}\biggr)+
p_1^i\biggl(\frac{\delta^{ij}}{r}+\frac{r^ir^j}{r^3}\biggr)\biggl(\frac{\delta^{jk}}{r_2}+\frac{r_2^jr_2^k}{r_2^3}\biggr)\,\biggr] P^k\nonumber\\
&&+\,\frac{1}{4}\,\biggl(\biggl[\frac{Z^2}{r_1^4}+\frac{Z^2}{r_2^4}\biggr]_\epsilon-2\,\frac{Z^2\,\vec{r}_1\cdot\vec{r}_2}{r_1^3r_2^3}\biggr)
+\frac{Z^3}{2}\biggl[\,\pi\delta^3(r_1)+\pi\delta^3(r_2)\,\biggr]\nonumber\\
&&+\,\frac{Z^2}{8}\biggl[\,p_1^i\,\frac{1}{r_1^2}\biggl(\delta^{ij}+3\,\frac{r_1^ir_1^j}{r_1^2}\biggr)\,p_1^j
+p_2^i\,\frac{1}{r_2^2}\biggl(\delta^{ij}+3\,\frac{r_2^ir_2^j}{r_2^2}\biggr)\,p_2^j\nonumber\\
&&
+\,2\,p_1^i\,\biggl(\frac{\delta^{ij}}{r_1}+\frac{r_1^ir_1^j}{r_1^3}\biggr)\biggl(\frac{\delta^{jk}}{r_2}
+\frac{r_2^jr_2^k}{r_2^3}\biggr)\,p_2^k\,\biggr]\biggr\rangle\,.\label{B30}
\end{eqnarray}

Finally, we calculate the correction due to the operator $H_7^M=H_{7a}^M+H_{7c}^M+H_{7d}^M$.
We split it into three parts, $B_7^M=B_{7a}^M+B_{7c}^M+B_{7d}^M$. The operator $H_{7a}^M$ reads
\begin{eqnarray}\label{B31}
H_{7a}^M&=&-\frac{1}{8}\,\biggr\{\bigl[p_1^i,V\bigr]\biggl(\frac{r^i r^j}{r}-3\,\delta^{ij}r\biggr)\bigl[V,p_2^j\bigr]+\bigl[p_1^i,V\bigr]
\biggl[\frac{p_2^2}{2},\,\frac{r^ir^j}{r}-3\,\delta^{ij}r\biggr]p_2^j\nonumber\\
&&+\,p_1^i\biggl[\frac{r^ir^j}{r}-3\,\delta^{ij}r,\,\frac{p_1^2}{2}\biggr]\bigl[V,p_2^j\bigr]
+p_1^i\biggl[\frac{p_2^2}{2},\,\biggl[\frac{r^ir^j}{r}-3\,\delta^{ij}r,\,\frac{p_1^2}{2}\biggr]\biggr]p_2^j\,\biggr\}\,.
\end{eqnarray}
The recoil correction due to this operator is
\begin{eqnarray}\label{B32}
\delta_M B_{7a}&=&\delta_M\,\biggl\langle-\,\frac{1}{8}\frac{Zr_1^i}{r_1^3}\frac{Zr_2^j}{r_2^3}\left(\frac{r^ir^j}{r}-3\delta^{ij}r\right)
+\frac{1}{4}\left(\frac{Z\vec{r}_1}{r_1^3}-\frac{Z\vec{r}_2}{r_2^3}\right)\cdot\frac{\vec{r}}{r^2}-\frac{1}{4}\biggl[\frac{1}{r}\biggr]^3_\epsilon\\
&&-\,\frac{Z}{8}\biggl[\,\frac{r_1^i}{r_1^3}\,p_2^k\left(\delta^{jk}\frac{r^i}{r}-\delta^{ik}\frac{r^j}{r}-\delta^{ij}\frac{r^k}{r}-\frac{r^ir^jr^k}{r^3}\right)p_2^j
+(1\leftrightarrow 2)\,\biggr]\nonumber\\
&&+\,\frac{1}{8}\biggl[\,p_2^j\,\frac{1}{r^4}\left(\delta^{jk}r^2-3r^jr^k\right)p_2^k+(1\leftrightarrow 2)\,\biggr]+\frac{1}{4}\biggl[\frac{1}{r^4}\biggr]_\epsilon
+\pi\,\delta^3(r)\nonumber\\
&&+\,\frac{1}{8}\,p_1^k\,p_2^l\biggl[-\frac{\delta^{il}\delta^{jk}}{r}+\frac{\delta^{ik}\delta^{jl}}{r}-\frac{\delta^{ij}\delta^{kl}}{r}-\frac{\delta^{jl}r^ir^k}{r^3}
-\frac{\delta^{ik}r^jr^l}{r^3}+3\,\frac{r^ir^jr^kr^l}{r^5}\,\biggr]p_1^i\,p_2^j\biggr\rangle.\nonumber
\end{eqnarray}
The operator $H_{7c}^M$ is
\begin{equation}\label{B33}
H_{7c}^M=-\,\frac{(d-1)}{16}\left[p_2^2,\left[p_1^2,\biggl[\frac{1}{r}\biggr]_\epsilon\right]\right]
\end{equation}
and the corresponding recoil correction is simply
\begin{eqnarray}\label{B34}
\delta_M B_{7c}&=&\delta_M\,\biggl\langle-\,\frac{(d-1)}{16}\left[p_2^2,\left[p_1^2,\biggl[\frac{1}{r}\biggr]_\epsilon\right]\right]\biggr\rangle\,.
\end{eqnarray}
Finally, the operator $H_{7d}^M$ is
\begin{eqnarray}
H_{7d}^M&=&i\,\frac{Z^2}{8M}\sum_{a,b}\frac{r_a^i}{r_a^3}\biggl[H-E,\,\frac{r_b^ir_b^j-3\,\delta^{ij}r_b^2}{r_b}\,p_b^j\biggr]\nonumber\\
&=&i\,\frac{Z^2}{8M}\sum_{a,b}\frac{r_a^i}{r_a^3}\biggl\{\bigl[V,p_b^j\bigr]\,\frac{r_b^ir_b^j-3\,\delta^{ij}r_b^2}{r_b}
+\biggl[\frac{p_b^2}{2},\,\frac{r_b^ir_b^j-3\,\delta^{ij}r_b^2}{r_b}\biggr]\,p_b^j\,\biggr\}\,.\label{B35}
\end{eqnarray}
The expectation value of this can then be written as
\begin{equation}\label{B36}
\delta_M B_{7d}= W_1+W_2
\end{equation}
where
\begin{eqnarray}
W_1&=&\biggl\langle-\,\frac{Z^2}{8}\sum_{a,b,c\neq b}\frac{r_a^i}{r_a^3}\left(\frac{Zr_b^j}{r_b^3}-\frac{r_{bc}^j}{r_{bc}^3}\right)\frac{r_b^ir_b^j-3\delta^{ij}r_b^2}{r_b}
-\frac{7\,Z^3}{4}\pi\delta^3(r_b)\biggr\rangle\nonumber\\
&=&\biggl\langle\frac{1}{4}\biggl[\frac{Z}{r_1}\biggr]^3_\epsilon+\frac14\biggl[\frac{Z}{r_2}\biggr]^3_\epsilon
+\frac{Z^3\,\vec{r}_1\cdot\vec{r}_2}{4r_1^3r_2^2}+\frac{Z^3\,\vec{r}_1\cdot\vec{r}_2}{4r_1^2r_2^3}
-\frac{7\,Z^3}{4}[\pi\delta^3(r_1)+\pi\delta^3(r_2)]\nonumber\\
&&+\,\frac{Z^2}{8}\sum_{b,c\neq b}\left(\frac{r_1^i}{r_1^3}+\frac{r_2^i}{r_2^3}\right)\frac{r_b^ir_b^j-3\delta^{ij}r_b^2}{r_b}\frac{r_{bc}^j}{r_{bc}^3}\biggr\rangle\,,\label{B37}
\end{eqnarray}
and
\begin{eqnarray}\label{B38}
W_2&=&\biggl\langle i\,\frac{Z^2}{16}\biggl(\,\sum_{a\neq b}\frac{r_a^i}{r_a^3}\,\biggl[p_b^2,\,\frac{r_b^ir_b^j-3\,\delta^{ij}r_b^2}{r_b}\biggr]\,p_b^j
+\sum_{b}\frac{r_b^i}{r_b^3}\,\biggl[p_b^2,\,\frac{r_b^ir_b^j-3\,\delta^{ij}r_b^2}{r_b}\biggr]p_b^j\biggr)\biggr\rangle\\
&=&\biggl\langle\frac{Z^2}{8}\,\sum_{a\neq b}p_b^k\,\frac{r_a^i}{r_a^3}\left(-\delta^{ik}\frac{r_b^j}{r_b}
+\delta^{jk}\frac{r_b^i}{r_b}-\delta^{ij}\frac{r_b^k}{r_b}-\frac{r_b^ir_b^jr_b^k}{r_b^3}\right)p_b^j\nonumber\\
&&+\,\frac18\biggl[\frac{Z^2}{r_1^4}+\frac{Z^2}{r_2^4}\biggr]_\epsilon+\frac{3\,Z^3}{4}[\pi\delta^3(r_1)+\pi\delta^3(r_2)]
+\frac{Z^2}{8}\sum_b p_b^j\,\frac{1}{r_b^4}\,\bigl(\delta^{jk}r_b^2-3r_b^jr_b^k\bigr)\,p_b^k\biggr\rangle\,.\nonumber
\end{eqnarray}
Summing all of the recoils parts $\delta_M B_i$ we get the result in Eq. (\ref{46}).

\begin{table*}
\caption{Expectation values of operators $Q_i$ with $i=1\ldots30$ for the $1^1S_0$, $2^1S_0$ and $2^1 P_1$ states.}
\scriptsize

\label{oprsQ}
\begin{ruledtabular}
\begin{tabular}{lddd}
 & 1^1S_0&2^1S_0&2^1P_1\\ \hline
$Q_1 =4 \pi \delta^3 (r_1)$   				     &  22.750526&16.455169&16.014493 \\
$Q_2 =4 \pi \delta^3 (r)$               	             &   1.336375&0.108679&0.009238\\
$Q_3 =4 \pi \delta^3(r_1)/r_2$                  	     &   33.440565&5.593743&3.934081\\	
$Q_4 =4 \pi \delta^3(r_1)\, p_2^2$ 	                     &   49.160046&7.578158&3.866237\\
$Q_5 =4 \pi \delta^3(r)/r_1$				     &   5.019713&0.440864&0.012785\\
$Q_6 =4 \pi\,\delta^3(r)\,P^2 $				     &   18.859765&1.800294&0.070787\\
$Q_7 =1/r$						     &   0.945818&0.249683&0.245024\\
$Q_8 =1/r^2$						     &   1.464771&0.143725&0.085798\\
$Q_9 =1/r^3$                    	                     &   0.989274&0.067947&0.042405\\
$Q_{10}=1/r^4$                  	                     &  -3.336384&-0.312402&0.008956\\
$Q_{11}=1/r_1^2$                	                     &  6.017409&4.146939&4.043035\\
$Q_{12}=1/(r_1 r_2)$            	                     &   2.708655&0.561861&0.491245\\
$Q_{13}=1/(r_1 r)$              	                     &   1.920944&0.340634&0.285360\\
$Q_{14}=1/(r_1 r_2 r)$          	                     &   4.167175&0.398366&0.159885\\
$Q_{15}=1/(r_1^2 r_2)$					     &   9.172094&1.472014&1.063079\\
$Q_{16}=1/(r_1^2 r)$					     &   8.003454&1.348761&1.002157\\
$Q_{17}=1/(r_1 r^2)$   					     &   3.788791&0.337891&0.105081\\
$Q_{18}=(\vec{r}_1\cdot\vec r)/(r_1^3 r^3)$                  &   3.270472&0.278353&0.010472\\
$Q_{19}=(\vec{r}_1\cdot\vec r)/(r_1^3 r^2)$                  &   1.827027&0.159078&0.043524\\
$Q_{20}=r_1^i r_2^j(r^i r^j-3\delta^{ij}r^2)/(r_1^3 r_2^3 r)$&   0.784425&0.063677&-0.004747\\
$Q_{21}=p_2^2/r_1^2$					     &   14.111960&2.064285&1.127058\\
$Q_{22}=\vec{p}_1/r_1^2\, \vec{p}_1$			     &  21.833598&16.459209&16.067214\\
$Q_{23}=\vec{p}_1/r^2\, \vec{p}_1$			     &   4.571652&0.499768&0.190797\\
$Q_{24}=p_1^i\,(r^i r^j+\delta^{ij} r^2)/(r_1 r^3)\, p_2^j$  &   0.811933&0.065354&0.053432\\
$Q_{25}=P^i\, (3 r^i r^j-\delta^{ij} r^2)/r^5\, P^j$	     &   -3.765488&-0.252967&0.013743\\
$Q_{26}=p_2^k \,r_1^i\,/r_1^3 (\delta^{jk} r^i/r - \delta^{ik} r^j/r
-\delta^{ij} r^k/r -r^i r^j r^k/r^3)\, p_2^j$		     &  -0.266894&-0.038928&-0.039976\\
$Q_{27}=p_1^2\, p_2^2$					     &   7.133710&1.428213&0.973055\\
$Q_{28}=p_1^2\,/r_1\, p_2^2$				     &   37.010643&5.955767&3.102248\\
$Q_{29}=\vec{p}_1\times\vec{p}_2\,/r\,\vec{p}_1\times\vec{p}_2$
							     &   4.004703&0.638960&0.216869\\
$Q_{30}=p_1^k \,p_2^l\,(-\delta^{jl} r^i r^k/r^3 - \delta^{ik} r^j r^l/r^3
+3r^i r^j r^k r^l/r^5)\, p_1^i\, p_2^j$			     &  -1.591864&-0.252663&-0.126416\\
\end{tabular}
\end{ruledtabular}
\end{table*}

\begin{table*}
\caption{Expectation values of operators $Q_i$ with $i=31\ldots50$, nonrelativistic energy $E$, the expectation value
of the Breit Hamiltonian $E^{(4)}$ and the first-order corrections $\delta_M E$ and $\delta_M E^{(4)}$
 for the $1^1S_0$, $2^1S_0$ and $2^1 P_1$ states.}
\scriptsize

\label{oprsQ2}
\begin{ruledtabular}
\begin{tabular}{lddd}
 & 1^1S_0&2^1S_0&2^1P_1\\ \hline
$Q_{31}=4 \pi \delta^3(r_1)\, \vec{p}_1\cdot\vec{p}_2$       &   5.610577&0.485629&0.281360\\
$Q_{32}=(\vec{r}_1\cdot\vec{r}_2)/(r_1^3 r_2^3)$             &  -0.683465&-0.054344&0.005113\\
$Q_{33}=\vec{p}_1\cdot\vec{p}_2$			     &   0.159069&0.009504&0.046045\\
$Q_{34}=\vec{P}\,/r_1\,\vec{P}$				     &   10.586465&5.103771&4.890226\\
$Q_{35}=\vec{P}\,/r\,\vec{P}$				     &   7.020556&1.367497&1.129114\\
$Q_{36}=\vec{P}\,/r_1^2\,\vec{P}$                            &  38.918728&18.764418&17.426840\\
$Q_{37}=\vec{P}\,/(r_1 r_2)\,\vec{P}$			     &  17.360500&3.093110&2.275085\\
$Q_{38}=\vec{P}\,/(r_1 r)\,\vec{P}$			     &  14.417322&2.139854&1.339969\\
$Q_{39}=\vec{P}\,/r^2\,\vec{P}$				     &  13.995389&1.425735&0.444219\\
$Q_{40}=p_1^2\,p_2^2\,P^2$				     &  244.833024&39.737868&20.202142\\
$Q_{41}=P^2\,p_1^i\, (r^i r^j + \delta^{ij} r^2)/r^3 \, p_2^j$
							     &  12.204592&1.693435&0.490552\\
$Q_{42}=p_1^i\,(r_1^i r_1^j + \delta^{ij} r_1^2)/r_1^4 \,  P^j$
						             & 45.454198&33.063647&32.258198\\
$Q_{43}=p_1^i\,(r_1^i r_1^j + \delta^{ij} r_1^2)/(r_1^3 r_2)\,  P^j$
						             &  16.864462&3.053603&2.163635\\
$Q_{44}=p_1^i\,p_2^k\,(r_1^ir_1^j+\delta^{ij}r_1^2)/r_1^3\,p_2^k\, P^j$
							     &  26.906923&4.533118&2.283665\\
$Q_{45}=p_2^i(r^i r^j+\delta^{ij} r^2)(r_1^jr_1^k
+\delta^{jk} r_1^2)/(r_1^3 r^3)\,  P^k$	 	    	     &  12.589902&1.471046&0.550295\\
$Q_{46}=p_1^i(r_1^i r_1^j+\delta^{ij} r_1^2)(r_2^jr_2^k
+\delta^{jk} r_2^2)/(r_1^3 r_2^3)\, p_2^k$		     &  1.225423&0.096713&0.111613\\
$Q_{47}=(\vec{r}_1\cdot\vec{r}_2)/(r_1^3 r_2^2)$             &  -0.275868&-0.021822&0.001588\\
$Q_{48}=r_1^i r^j(r_1^i r_1^j-3\delta^{ij} r_1^2)/(r_1^4r^3)$&  -2.285118&-0.185238&-0.034770\\
$Q_{49}=r_1^ir^j(r_2^ir_2^j-3\delta^{ij}r_2^2)/(r_1^3r_2r^3)$&  -3.574722&-0.306798&-0.074979\\
$Q_{50}=p_2^k\,r_1^i/r_1^3\,(\delta^{jk}r_2^i/r_2-\delta^{ik}r_2^j/r_2
-\delta^{ij}r_2^k/r_2-r_2^ir_2^jr_2^k/r_2^3)\,p_2^j$	     &  -0.071814&0.014329&0.041860\\
$E$ &-2.903724377 & -2.145974046& -2.123843086\\
$E^{(4)}$ &-1.951754768 &-2.034167340 &-2.040025575 \\
$\delta_M E$ &3.062793852 &2.155477910 &2.169887611 \\
$\delta_M E^{(4)}$ &-2.159371705 &-0.069625849 & -0.058484955\\
\end{tabular}
\end{ruledtabular}
\end{table*}

\begin{table*}
\caption{Expectation values of operators $\delta_M\langle Q_i\rangle$ with $i=1\ldots30$ for the $1^1S_0$, $2^1S_0$ and $2^1 P_1$ states.}
\scriptsize

\label{oprsdQ}
\begin{ruledtabular}
\begin{tabular}{lddd}
 & 1^1S_0&2^1S_0&2^1P_1\\ \hline
$\delta_M\langle Q_1 \rangle $   				     &  -69.398419&-49.370647&-47.548301\\
$\delta_M\langle Q_2 \rangle $               	        	     &   -4.164065&-0.303860&-0.071149\\
$\delta_M\langle Q_3 \rangle $                  		     &  -140.863781&-22.886485&-17.954636\\
$\delta_M\langle Q_4 \rangle $ 	                		     &  -264.235067&-39.376218&-23.782626\\
$\delta_M\langle Q_5 \rangle $					     &  -21.752541&-1.810273&-0.115562\\
$\delta_M\langle Q_6 \rangle $					     &  -104.659635&-9.811620&-0.734559\\
$\delta_M\langle Q_7 \rangle $					     &  -0.884405&-0.254546&-0.394523\\
$\delta_M\langle Q_8 \rangle $					     &  -2.818398&-0.266907&-0.305911\\
$\delta_M\langle Q_9 \rangle $        	                  	     &  -1.015798&-0.042815&-0.216329\\
$\delta_M\langle Q_{10} \rangle$          	              	     & 14.670321&1.241088&-0.016021\\
$\delta_M\langle Q_{11} \rangle$      	                 	     & -12.344317&-8.297087&-8.038384\\
$\delta_M\langle Q_{12} \rangle$        	                     &   -5.755090&-1.156557&-1.274719\\
$\delta_M\langle Q_{13} \rangle$              	                     &  -3.923779&-0.687748&-0.772874\\
$\delta_M\langle Q_{14} \rangle$          	               	     &  -13.208243&-1.217078&-0.704072\\
$\delta_M\langle Q_{15} \rangle$				     &   -29.209816&-4.532140&-3.865798\\
$\delta_M\langle Q_{16} \rangle$				     &   -25.139317&-4.116908&-3.618037\\
$\delta_M\langle Q_{17} \rangle$   				     &  -11.755788&-0.997079&-0.498120\\
$\delta_M\langle Q_{18} \rangle$   		                     &  -14.692291&-1.220964&-0.076044\\
$\delta_M\langle Q_{19} \rangle$                   		     &  -6.384958&-0.549039&-0.222341\\
$\delta_M\langle Q_{20} \rangle$  			             &  -5.471095&-0.509842&-0.028997\\
$\delta_M\langle Q_{21} \rangle$				     &   -61.053735&-8.609657&-5.848982\\
$\delta_M\langle Q_{22} \rangle$				     & -89.811452&-65.992539&-64.011907\\
$\delta_M\langle Q_{23} \rangle$				     &  -19.418528&-2.078930&-1.071679\\
$\delta_M\langle Q_{24} \rangle$				     &  -6.349789&-0.818061&-0.508001\\
$\delta_M\langle Q_{25} \rangle$	     		 	     &   20.318585&1.280443&-0.069997\\
$\delta_M\langle Q_{26} \rangle$		  		     &  0.019487&0.013046&0.262948\\
$\delta_M\langle Q_{27} \rangle$				     &   -31.111811&-5.980380&-5.023306\\
$\delta_M\langle Q_{28} \rangle$				     &   -199.698515&-31.075150&-19.296491\\
$\delta_M\langle Q_{29} \rangle$  				     &  -21.211342&-3.263956&-1.458861\\
$\delta_M\langle Q_{30} \rangle$				     &  9.913115&1.535897&0.868894\\
\end{tabular}
\end{ruledtabular}
\end{table*}

\begin{table*}
\caption{Results for the $\alpha^6\,m^2/M$ contribution to ionization energies of the $1^1S_0$, $2^1S_0$ and $2^1P_1$
states of helium.}

\label{Es}
\begin{ruledtabular}
\begin{tabular}{lddd}
$\alpha^6\,m^2/M$ & \multicolumn{1}{c}{$1^1S_0$} & \multicolumn{1}{c}{$2^1S_0$} & \multicolumn{1}{c}{$2^1P_1$} \\ \hline
$E_\textrm{i}$      &  -2.676\,12(3) &  -0.245\,21     &  -0.482\,76(12) \\ 
$E_\textrm{iii}$    &   7.337\,46    &   0.680\,17     &  -3.419\,39   \\   
$E_\textrm{v}$      & -48.911\,81    & -52.988\,42     & -53.940\,11   \\
$E_\textrm{vi}$     &  60.445\,89    &  53.983\,65     &  54.110\,62   \\ \hline
Subtotal          & 16.195\,42(3)     &1.430\,19       & -3.731\,64(12)\\
$E_\textrm{vii}$    &-152.161\,17    &  -9.857\,35     &   2.630\,40   \\ \hline
$\delta_M E^{(6)}$  &-135.965\,75(3) &  -8.427\,16     &  -1.101\,24(12) \\
$\delta_M E^{(6)}(\textrm{kHz}\cdot h)$& -347.79 & -21.56 & -2.82 \\
\end{tabular}
\end{ruledtabular}
\end{table*}

\begin{table*}
\caption{Breakdown of theoretical contributions to the $2^1S$--$2^3S$ centroid transition frequencies in
${}^4\textrm{He}$, in MHz. }
\label{tab:abs}
\begin{center}
\begin{tabular}{c | w{11.6}w{8.6}w{4.5}w{4.5}w{10.8}}
\hline
\hline
   & \centt{$(m/M)^0$} & \centt{$(m/M)^1$}  & \centt{$(m/M)^2$} & \centt{$(m/M)^3$}   &  \centt{Sum} \\
\hline
 $\alpha^2$        &   192\,490\,838.755 &  -24\,529.467 &       -6.511 &      0.004    &   192\,466\,302.781 \\
 $\alpha^4$        &         45\,657.859 &        -7.628 &        0.003 &  \textrm{---} &         45\,650.234    \\
 $\alpha^5$        &         -1\,243.670 &         0.173 & \textrm{---} &  \textrm{---} &         -1\,243.497  \\
 $\alpha^6$        &              -6.947 &         0.008 & \textrm{---} &  \textrm{---} &              -6.939 \\
 $\alpha^7$        &            1.4(0.7) &  \textrm{---} & \textrm{---} &  \textrm{---} &               1.4(0.7) \\
 FNS               &              -0.607 &  \textrm{---} & \textrm{---} &  \textrm{---} &              -0.607 \\
\hline
Total                    &&&&&    192\,510\,703.4(0.7)\\
Exp. \cite{rooij:11}     &&&&&    192\,510\,702.145\,6(1\,8) \\
 \hline \hline
\end{tabular}
\end{center}
\end{table*}

\begin{table*}
\caption{Breakdown of theoretical contributions to the ${}^3\textrm{He}-{}^4\textrm{He}$ isotope shift of
the $2^1S$--$2^3S$ centroid transition frequencies, for the point nucleus, in kHz. EMIX is the contribution due to
the mixing of the $2^1S$ and $2^3S$ states that comes form the contact Fermii interaction.
The related uncertainty of $\alpha^6\,(m/M)^2$ term due to hyperfine mixing with other states is estimated
to be of 0.15 kHz.}
\label{tab:is}
\begin{center}
\begin{tabular}{c | w{10.6}w{6.6}w{6.6}w{10.6}}
\hline
\hline
 $m$  & \centt{$(m/M)^1$}  & \centt{$(m/M)^2$} & \centt{$(m/M)^3$}   &  \centt{Sum} \\
\hline
 $\alpha^2$          &  -8\,026\,758.52 &     -4\,958.33 & 5.07         & -8\,031\,711.78 \\
 $\alpha^4$          &       -2\,496.23 &           2.08 & \textrm{---} &      -2\,494.15 \\
 $\alpha^5$          &            56.61 & \textrm{---}   & \textrm{---} &           56.61 \\
 $\alpha^6$          &             2.73 & 0.00(15)       & \textrm{---} &            2.73(15) \\
 $\alpha^7$          &        -0.21(11) &                &              &             -0.21(11)  \\
 NPOL \cite{rooij:11}&             0.20(2)& \textrm{---} & \textrm{---} &           0.20(2) \\
 EMIX                &        \textrm{---}&  80.69       & \textrm{---} &        80.69 \\
\hline
Present theory       &&&&    -8\,034\,065.91(19) \\
 \hline \hline
\end{tabular}
\end{center}
\end{table*}

\begin{table*}
\caption{Determination of the nuclear charge difference $\delta r^2$ from the measurement by Rooij {\em et al.} in Ref.~\cite{rooij:11},
in kHz. \label{tab:rms:roij}
}
\begin{ruledtabular}
  \begin{tabular}{l.l}
$E(^3{\rm He},2^1S^{F=1/2} - 2^3S^{F=3/2}) - E(^4{\rm He},2^1S - 2^3S)$ &  -5\,787\,719x.2(2.4) &Exp. \cite{rooij:11}\\
$\delta E_{\rm hfs}(2^3S^{3/2})$& -2\,246\,567x.059(5) & Exp. \cite{schluesser:69,rosner:70}\\
$-\delta E_{\rm iso}(2^1S - 2^3S)$ (point nucleus) &8\,034\,065x.91\,(19) & Theory, Table~\ref{tab:is} \\[1ex]
$\delta E$                     &-220x.4(2.4) & \\
$C$                            &-214x.66\,(2)\,\,\, {\rm kHz/fm}^2 & Ref. \cite{heis} \\
$\delta r^2 = r^2(^3\mbox{\rm He}) - r^2(^4\mbox{\rm He})$                   & 1x.027\,(11)\;{\rm fm}^2             &
  \end{tabular}
\end{ruledtabular}
\end{table*}

\end{document}